\documentclass[10pt,a4paper,final]{iopart}

\usepackage{graphicx}
\usepackage{dcolumn,multirow}
\usepackage{bm}
\usepackage{iopams}
\usepackage{hyperref}
\hypersetup{colorlinks=true, citecolor=blue, urlcolor=blue, linkcolor=blue}
\usepackage{color}
\usepackage{epsfig}
\usepackage{epstopdf}

\usepackage{cancel}

\def\nuc#1#2{\relax\ifmmode{}^{#1}{\protect\text{#2}}\else${}^{#1}$#2\fi}

\definecolor{vert}{rgb}{0,0.6,0}

\begin{document}
  \newcommand {\nc} {\newcommand}
  \nc {\beq} {\begin{eqnarray}}
  \nc {\eeq} {\nonumber \end{eqnarray}}
  \nc {\eeqn}[1] {\label {#1} \end{eqnarray}}
  \nc {\eol} {\nonumber \\}
  \nc {\eoln}[1] {\label {#1} \\}
  \nc {\ve} [1] {\mbox{\boldmath $#1$}}
  \nc {\ves} [1] {\mbox{\boldmath ${\scriptstyle #1}$}}
  \nc {\mrm} [1] {\mathrm{#1}}
  \nc {\half} {\mbox{$\frac{1}{2}$}}
  \nc {\thal} {\mbox{$\frac{3}{2}$}}
  \nc {\fial} {\mbox{$\frac{5}{2}$}}
  \nc {\la} {\mbox{$\langle$}}
  \nc {\ra} {\mbox{$\rangle$}}
  \nc {\eq} [1] {(\ref{#1})}
  \nc {\Eq} [1] {Eq.~(\ref{#1})}
  \nc {\Ref} [1] {Ref.~\cite{#1}}
  \nc {\Refc} [2] {Refs.~\cite[#1]{#2}}
  \nc {\Sec} [1] {Sec.~\ref{#1}}
  \nc {\chap} [1] {Chapter~\ref{#1}}
  \nc {\anx} [1] {Appendix~\ref{#1}}
  \nc {\tbl} [1] {Table~\ref{#1}}
  \nc {\Fig} [1] {Fig.~\ref{#1}}
  \nc {\ex} [1] {$^{#1}$}
  \nc {\Sch} {Schr\"odinger }
  \nc {\flim} [2] {\mathop{\longrightarrow}\limits_{{#1}\rightarrow{#2}}}
  \nc {\IR} [1]{\textcolor{red}{#1}}
  \nc {\IG} [1]{\textcolor{vert}{#1}}
  \nc {\IB} [1]{\textcolor{blue}{#1}}
  \nc{\pderiv}[2]{\cfrac{\partial #1}{\partial #2}}
  \nc{\deriv}[2]{\cfrac{d#1}{d#2}}

\title[Extension of the ratio method to proton-rich nuclei]{Extension of the ratio method to proton-rich nuclei}


\author{X.~Y.~Yun$^{1}$, F.~Colomer$^{2,3}$, D.~Y.~Pang$^{1,4}$ and P.~Capel$^{2,3}$}
\address{$^1$ School of Physics and Nuclear Energy Engineering, Beihang University, Beijing 100191, China}
\address{$^2$ Physique Nucl\' eaire et Physique Quantique, Universit\'e Libre de Bruxelles (ULB), B-1050 Brussels}
\address{$^3$ Institut f\"ur Kernphysik, Johannes Gutenberg-Universit\"at Mainz, 55099 Mainz, Germany}
\address{$^4$ Beijing Key Laboratory of Advanced Nuclear Materials and Physics, Beihang University,	Beijing 100191, China}
\ead{yunxy@buaa.edu.cn, fcolomer@ulb.ac.be, dypang@buaa.edu.cn and pcapel@uni-mainz.de}

\date{\today}

\begin{abstract}
The ratio method has been developed to improve the study of one-neutron halo nuclei through reactions.
By taking the ratio of angular distributions for two processes, viz. breakup and elastic scattering, this new observable is nearly independent of the reaction mechanism and hence much more sensitive to the projectile structure than the cross sections for each single process.

We study the extension of the ratio method to proton-rich nuclei and also explore the optimum experimental conditions for measuring this new observable. 
We compare accurate dynamical calculations of reactions for proton-rich projectiles to the prediction of the ratio method. 
We use the dynamical eikonal approximation that provides good results for this kind of reactions at intermediate energy.

Our tests for $^8$B, an archetypical one-proton halo nucleus, on Pb, Ni, and C targets at 44~MeV/nucleon show that, the ratio works less well than for neutron halos due to the non-negligible Coulomb interaction between the valence proton and the target.
Nevertheless, thanks to its strong sensitivity to the single-particle structure of the projectile, the ratio method still provides pertinent information about nuclear structure on the proton-rich side of the valley of stability.
To account for the lower quality of the method applied to charged systems, we suggest variations in its application from the original idea.
Interestingly the method is not affected if energy ranges---or \textit{bins}---are considered in the projectile continuum.
This makes the ratio easier to measure experimentally by increasing the breakup cross section.
We also extend our analysis to $^{17}$F, $^{25}$Al, and $^{27}$P, whose study is of interest to both nuclear astrophysics and nuclear structure.

We show that, albeit less precise than for one-neutron halo nuclei, nuclear-structure information can be inferred from the ratio method applied to exotic proton-rich nuclei.
For nuclei in which the valence proton is deeply bound and/or sits in an $l\ge2$ orbital, the method provides only estimates of nuclear-structure features, like the one-proton separation energy or the orbital angular momentum of the valence proton in the ground state.
When the valence proton is loosely bound in an $s$ or $p$ orbital, viz. for proton halo nuclei, more detailed structure information can be obtained through this new reaction observable.

\end{abstract}

\noindent{\it{Keywords}}: {Halo nuclei, proton-rich nuclei, elastic scattering, breakup reaction, ratio of cross sections}

\submitto{\jpg}

\section{\label{sec:intro}Introduction}

The development of radioactive-ion beam (RIB) facilities in the mid 80's has opened the door to the study of nuclear structure away from the $\beta$-stability line. This technical breakthrough has enabled the discovery of new phenomena that are not observed in stable nuclei. The existence of neutron halo or neutron skin in some nuclei \cite{Tan85r, JRF04, KHH16}, the emergence of new magic numbers \cite{OKS00, KNP09, STA13,  RAA15}, and shape coexistence \cite{HW11,YWX16} are so many examples of these exotic features, which challenge the current nuclear theory.

Because they are located beyond the $\beta$-stability line, these exotic nuclei cannot be studied through usual spectroscopic methods.
Experimentalists must then rely on indirect techniques, such as reactions, to analyze their structure.
For example, breakup reactions are one of the mostly used tools to study halo nuclei \cite{Tan96}.
In these reactions the loosely bound valence nucleon dissociates from the core of the nucleus during its interaction with the target, hence revealing its strongly clusterized internal structure.
In order to extract valuable nuclear-structure information from experiment, an accurate reaction model coupled to a realistic description of the projectile is needed.
Various such models have been developed within the last thirty years (see, e.g., \Ref{BC12} for a recent review).
Breakup calculations depend on the optical potentials, which simulate the interaction between the target and the projectile internal clusters.
The resulting breakup cross sections may be quite uncertain due to the ambiguities inherited from optical potentials \cite{CGB04}.
Such problems make the study of cluster structures with breakup measurements more difficult than initially thought.

The ratio method was recently suggested to circumvent this issue in the study of one-neutron halo nuclei \cite{capel2011, capel2013}.
The core idea of this method is to look at the ratios of angular distributions for different reaction channels that occur in the collision, viz. breakup and elastic scattering.
Theoretically, it is based on the recoil excitation and breakup (REB) model of reactions induced by one-neutron halo nuclei \cite{johnson1997,RCJ1997}.
This model predicts this ratio to be independent of the reaction mechanism, and, in particular, to be insensitive to the optical potentials, whose influence on the different cross sections cancel out when taking their ratio.
According to the REB model, the ratio should be equal to a form factor that depends only on the projectile wave functions, making this new observable much more sensitive to its internal structure than the individual reaction cross sections.

The excellent results obtained for one-neutron halo nuclei, even beyond the range of validity of the REB \cite{colomer2016}, lead us to consider the extension of the ratio method to study the single-particle structure of proton-rich nuclei.
This extension is not self-evident because the REB is built on two simplifying assumptions that are likely to be breached for proton-rich nuclei.
First, the REB neglects the interaction between the valence nucleon and the target.
For a valence proton it is not fully clear that this can be done because it is always sensitive to the Coulomb field of the target.
Second, the adiabatic---or sudden---approximation is applied to the treatment of the projectile dynamics \cite{johnson1997,RCJ1997}.
This is usually valid for short-ranged nuclear interactions.
However due to the additional Coulomb force between the proton and the target, this approximation might lapse.
The goal of this work is to evaluate the significance of these approximations in reactions involving proton-rich nuclei and see whether the ratio method still holds for these exotic systems.
We also study the best conditions to explore this method experimentally.

After a brief description of the ratio method in \Sec{sec:theory}, we analyze in \Sec{sec:8B} the collision of $^8$B---an archetypical one-proton halo nucleus---on various targets at intermediate energy.
As in Refs.~\cite{capel2011,capel2013}, we consider for reaction model the dynamical eikonal approximation (DEA) \cite{baye2005,goldstein2006}, which has shown to provide excellent results for the breakup of $^8$B at intermediate energies \cite{GCB07}.
We estimate the validity of the ratio method by confronting these reaction calculations to the REB predictions.
Then, in \Sec{sec:Sensitivity}, we study in detail the sensitivity of the ratio to various aspects of the projectile structure: its binding energy, the asymptotic normalization constant (ANC) of its radial wave function, the partial-wave in which the halo proton is bound to the core, etc. 
We also check how the ratio behaves when considering a range of energies in the continuum instead of a single energy.
Based on these tests, we present various ways the ratio method can be applied to study the structure of proton-rich nuclei.
In \Sec{sec:otherhalo}, we explore the possibility to use the ratio method to study other proton-rich nuclei, namely, \ex{17}{F}, \ex{25}{Al}, and \ex{27}{P}, whose study is of interest to both nuclear astrophysics \cite{Togano-PRC-2011,Chen-PRC-2012, Jung-PRC-2012, Fortune-PRC-2015, Marganiec-PRC-2016} and nuclear structure 
\cite{Hag10,XuXX-PLB-2013}. 
These nuclei exhibit the clear single-particle structure of a proton outside a core, for which the ratio method could fit well.
We look in particular for the optimum experimental conditions, hoping to lay the ground for the experimental validation of the method with proton-rich nuclei.
Our conclusions are drawn in \Sec{sec:conclusion}.

\section{\label{sec:theory}The ratio method in a nutshell}

\subsection{\label{model}Model of reaction}

We describe the collision within a three-body model: a two-body projectile impinging on a one-body target.
The projectile $P$ is composed of a proton $p$ of spin $1/2$, which is loosely bound to a core $c$ of atomic and mass numbers $Z_c$ and $A_c$, respectively.
The atomic and mass numbers of the projectile are thus $Z_P=Z_c+1$ and \mbox{$A_P=A_c+1$}, respectively.
For simplicity, we neglect the spin and internal structure of the core.
Such a two-body structure is described by the Hamiltonian
\begin{equation}
H_0 = -\frac{\hbar^2}{2\mu_{cp}}\Delta_{\ve{r}} + V_{cp}(\ve{r}), \label{e1}
\end{equation}
where $\mu_{cp} = A_c m_N/A_P$ is the $c$-$p$ reduced mass, with $m_N$ the nucleon mass, $\ve{r}$ is the relative coordinate of the proton to the core and $V_{cp}$ is a phenomenological potential that simulates the core-proton interaction.

In this model, the states of the projectile are described by the eigenstates of $H_0$, which read, in the partial wave $ljm$,
\begin{equation}
H_0\;\phi_{ljm}(E,\ve{r}) = E\;\phi_{ljm}(E,\ve{r}),
\end{equation}
where $l$ is the orbital angular momentum for the $c$-$p$ relative motion, $j$ is the total angular momentum obtained from the coupling of $l$ with the proton spin, and $m$ is its projection.

Negative-energy states are discrete and form the bound spectrum of the nucleus.
To distinguish them, we introduce within their notation the number $n$ of nodes in the radial wave function.
The positive-energy eigenstates of $H_0$ describe the continuum of the projectile, i.e., the states in which the proton is dissociated from the core.
This continuum may include one-proton resonances.
The parameters of $V_{cp}$ are adjusted to reproduce the experimentally known low-energy states of the projectile, its bound states and, when possible, some of its resonances.

The target $T$ is considered as a structureless body of atomic and mass numbers $Z_T$ and $A_T$, respectively.
Its interactions with the core and the valence proton are simulated by the optical potentials $V_{cT}$ and $V_{pT}$, respectively.
These potentials are chosen from the literature, or built from folding procedures \cite{XP13} and reproduce the elastic scattering of each of the constituents of the projectile with the target.

If we define $\ve{R}$ as the $P$-$T$ relative coordinate, the core- and proton-target coordinates read $\ve{R}_{cT}= \ve{R} - \frac{1}{A_P}\ve{r}$ and $\ve{R}_{pT} = \ve{R} + \frac{A_c}{A_P}\ve{r}$, respectively.
In the Jacobi set of coordinates $\{\ve{r},\ve{R}\}$, the three-body Schr\"odinger equation that describes the collision reads
\begin{eqnarray}
\left[-\vphantom{\frac{\hbar^2}{2\mu_{PT}}}\right.\left.\frac{\hbar^2}{2\mu_{PT}}\Delta_{\bm{R}} + H_0 + V_{cT}(R_{cT}) + V_{pT}(R_{pT}) \right]&&\;\Psi(\bm{r},\bm{R}) \nonumber \\
&&= E_{\mathrm{tot}}\;\Psi(\bm{r},\bm{R}),\label{eq:Schr}
\end{eqnarray}
where $\mu_{PT}=A_P A_Tm_N/(A_P+A_T)$ is the $P$-$T$ reduced mass and $\Psi$ is the three-body wavefunction.
Initially, the projectile is in its ground state $n_0l_0j_0$ of energy $E_0$ and has an initial $P$-$T$ relative momentum $\hbar K_0$.
This fixes the total energy of the system in its center-of-mass rest frame to $E_{\mathrm{tot}}= \hbar^2K_0^2/2\mu_{PT} + E_0$.
With $\ve{\hat Z}$ the direction of the incoming beam, the initial condition reads
\begin{equation}
\Psi(\bm{r},\bm{R}) \mathop{\longrightarrow}\limits_{Z \rightarrow -\infty} e^{i\{K_0 Z+\eta \ln [K_0(R-Z)]\}}\;\phi_{n_0l_0j_0m_0}(E_0,\bm{r}),
\end{equation}
where $\eta =Z_T Z_P e^2/(4\pi\epsilon_0\hbar^2K_0/\mu_{PT})$ is the $P$-$T$ Sommerfeld parameter.

To solve \Eq{eq:Schr}, we choose to use the dynamical eikonal approximation (DEA) \cite{baye2005,goldstein2006}.
This approximation simplifies the equation to be solved, allows for shorter computational times, and is very accurate at intermediate energies \cite{CEN12}.
In particular, for the proton-rich nuclei studied here, the DEA provides excellent agreement with the MSU breakup experiments on $^8$B at 44, 81, and 83~MeV/nucleon \cite{GCB07,Dav98,Dav01l,Dav01c}.
In this work we use the Coulomb-corrected version of the DEA detailed in Ref.~\cite{fukui2014}.

\subsection{The ratio idea}
In \Ref{capel2011}, a new reaction observable has been suggested to study one-neutron halo nuclei.
Instead of looking at elastic scattering or breakup cross sections separately, the idea is to measure the ratio of cross sections, and more precisely the ratio of the breakup angular distribution for a given energy $E$ in the core-valence-nucleon continuum ($d\sigma_{\rm BU}/dEd\Omega$) and the so-called summed cross section, which corresponds to all the quasi-elastic processes: elastic and inelastic scattering, and breakup
\begin{eqnarray}\label{eq:Xsum}
\frac{d\sigma_{\mathrm{sum}}}{d\Omega}&=&\frac{d\sigma_{\mathrm{el}}}{d\Omega}+\frac{d\sigma_{\mathrm{inel}}}{d\Omega}+\int\frac{d\sigma_{\rm BU}}{dEd\Omega}dE.
\end{eqnarray}
The ratio observable hence reads
\begin{equation}
\mathcal{R}_{\mathrm{sum}}(E,\bm{Q}) = \frac{d\sigma_{\mathrm{BU}}/dEd\Omega}{d\sigma_{\mathrm{sum}}/d\Omega},
\label{eq:ratio}
\end{equation}
where
\begin{equation}
\bm{Q}= \frac{1}{A_P}(K_0 \ve{\hat Z}-\bm{K}')
\label{eq:Q}
\end{equation}
is proportional to the transferred momentum from the initial $\hbar K_0 \ve{\hat Z}$ to the final $\hbar\ve{K}'$ momenta, and is approximately related to the scattering angle between the projectile center of mass and the target after the collision by $Q\simeq \frac{2}{A_P}K_0\sin(\theta/2)$.

The recoil excitation and breakup model (REB) \cite{johnson1997,RCJ1997} predicts this ratio to be equal to a form factor that depends only on the structure of the projectile
\begin{equation}
\mathcal{R}_{\mathrm{sum}}(E,\bm{Q}) \mathop{=}\limits^{\mathrm{REB}} |F_{E,0}(\bm{Q})|^2,
\label{eq:REBratio}
\end{equation}
with
\begin{eqnarray}\label{eq:formf}
|F&_{E,0}(\bm{Q})|^2 = \nonumber \\
&\frac{1}{2j_0 +1}\sum_{m_0}\sum_{ljm}\left|\int\phi_{ljm}(E,\bm{r})\phi_{n_0l_0j_0m_0}(E_0,\bm{r})e^{i\bm{Q}\cdot\bm{r}}d\bm{r}\right|^2.
\end{eqnarray}
This observable should thus be independent of the reaction mechanism and therefore provide a very accurate probe of the nuclear structure.
In particular it should be independent of the optical potentials chosen to simulate the interaction between the projectile constituents and the target, which can significantly affect reaction calculations \cite{CGB04}.
Note that this form factor differs from the $dB({\rm E1})/dE$ strength from the ground state to the $c$-$p$ continuum, but at (very) small $Q$, for which $e^{i\bm{Q}\cdot\bm{r}}\sim1+i\,\bm{Q}\cdot\bm{r}$.

The REB is based on two simplifying approximations, which enable the exact resolution of \Eq{eq:Schr}.
First, it neglects the interaction between the valence nucleon and the target [viz.\ $V_{pT}=0$ in \Eq{eq:Schr}] and, second, it assumes the adiabatic---or sudden---approximation (viz.\ $H_0 \simeq E_0$).
In Refs.~\cite{capel2011,capel2013}, it has been shown using the DEA that the ratio idea derived from the REB works well for one-neutron halo nuclei even if these two conditions are not met in real cases.
In a later work, the extension of the ratio method to low beam energy, viz.\ down to 20~MeV/nucleon, has been demonstrated \cite{colomer2016}.

One practical conclusion of these previous analyzes is that although the ratio should be independent of the reaction process, it seems more efficient in collisions where the Coulomb interaction is less significant, i.e. on light targets, because in these cases, the adiabatic approximation is better justified.
This suggests that the ratio method could also be used to study the single-particle structure of proton-rich nuclei, such as proton-halo nuclei, even though the presence of a Coulomb term in the interaction between the valence proton and the target certainly breaches the $V_{pT}=0$ hypothesis made within the REB and would make the adiabatic approximation less valid.
The main goal of the present work is to see whether the ratio method can be extended to the case of proton-rich nuclei and what are the best experimental conditions to measure it in practice.
We proceed as in the previous works \cite{capel2011,capel2013,colomer2016} and compare the REB prediction \eq{eq:REBratio} to precise DEA reaction calculations for this observable.
To this aim, we initiate our study with $^8$B, which exhibits a very clear one-proton halo structure, before extending the idea to other proton-rich nuclei.

\section{\label{sec:8B}Extension of the ratio method to a $^{8}$B projectile}

\subsection{Inputs to the reaction model}\label{inputs}

\subsubsection{Description of $^8$B}\label{8B}
The nucleus $^8$B has a strong $^7$Be-$p$ cluster structure and is usually considered as the archetypical one-proton halo nucleus. It therefore constitutes the ideal test case to study the extension of the ratio method to proton-rich nuclei. In this section, we study the collision of $^8$B on Pb, C, and Ni targets at 44 MeV/nucleon. The spectrum of $^8$B includes only one bound state, which exhibits a one-proton separation energy $S_p$ of a mere 137 keV. Its $2^+$ spin and parity are obtained predominantly from the coupling of a $0p_{3/2}$ proton with the $\frac{3}{2}^-$ spin of the ground state of $^7$Be \cite{BDT94}. Following Refs.~\cite{GCB07,MTT02}, we use the simplified version of the description of $^8$B developed by Esbensen and Bertsch in \Ref{esbensen1996}. This description neglects the spin of the core and reproduces the bound state of $^8$B as a $0p_{3/2}$ proton bound to a spinless $^7$Be core.

\subsubsection{Optical potentials $V_{cT}$ and $V_{pT}$}\label{optpo}
The $^{7}$Be core being the mirror nucleus of $^{7}$Li, we follow Refs.~\cite{GCB07,MTT02} and choose to simulate the $c$-$T$ interaction by optical potentials that were fitted to $^7$Li elastic-scattering data.
For the $^{208}$Pb target, we extrapolate the global potential suggested by Cook in \Ref{cook1982} to reproduce the elastic scattering of $^7$Li on various targets, from $^{24}$Mg to $^{208}$Pb, in an energy range 28--88~MeV.
For the $^{12}$C target, we consider the potential developed in \Ref{nadasen1995} to fit elastic-scattering data of $^7$Li off $^{12}$C at 350~MeV. For the $^{58}$Ni target instead, we rescale the potential developed in \Ref{CLARK1995416} to fit elastic-scattering data of $^4$He off $^{58}$Ni at 240~MeV.

To test the independence of the ratio to the $c$-$T$ interaction, we also use,  for the Pb and C targets, the potentials from Refs.~\cite{chuev1971,schumacher1973}, which are listed in the Perey and Perey compilation \cite{perey1976}.
The former has been fitted to reproduce the elastic scattering of $^{6}$Li on $^{208}$Pb at 30~MeV, whereas the latter simulates the scattering of $^{7}$Li off $^{12}$C at 36~MeV.  
We have rescaled the radius of the former to account for the mass difference between both projectiles.
Although these potentials have been developed for energies well below the ones considered here, we neglect the possible energy dependence of these interactions.
This second set of potentials will be referred to as $V'_{cT}$ in the following.

To simulate the $p$-$T$ interaction, we use the nucleon-target global optical potential of Becchetti and Greenlees \cite{becchetti1969} on the $^{208}$Pb target.
For the $^{12}$C and $^{58}$Ni targets, we consider the Koning-Delaroche global parametrization \cite{koning2003}.

\subsubsection{Numerical conditions of the calculations}\label{numerics}
The cross sections entering the computation of $\mathcal{R}_{\mathrm{sum}}$ [see \Eq{eq:ratio}] are  calculated within the DEA \cite{baye2005,goldstein2006} using the Coulomb correction from Ref.~\cite{fukui2014}.
The computations are done with the algorithm presented in Ref.~\cite{capel2003}, which expands the projectile wave-function over a mesh on the unit sphere containing $N_{\theta}\times N_{\phi}$ points.
At 44~MeV/nucleon, we go up to $14\times27$ points for the $^{12}$C and $^{58}$Ni targets and $10\times19$ points for the $^{208}$Pb target.
The radial mesh is quasi-uniform, contains $N_r=800$ points, and extends up to $r_{N_r}=800$~fm.
The impact parameters considered in the calculations are discretized in steps $h_b=0.25$--5~fm in the range of $b=0$--200~fm for all targets.

\subsection{$^{208}$Pb target at 44 MeV/nucleon}\label{8BPb}

\begin{figure}[htb]
	\centering
	\includegraphics[width=0.7\columnwidth]{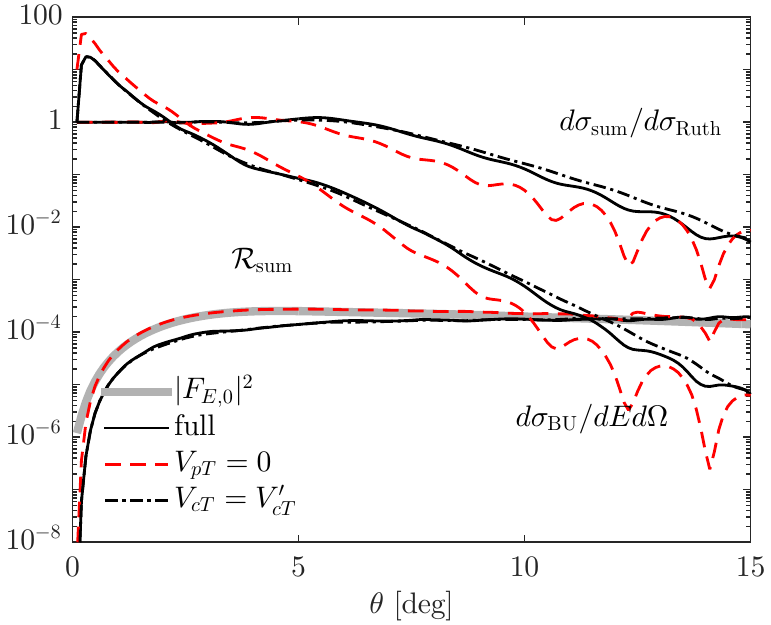}
	\caption{\label{f1}
Analysis of the ratio method for $^{8}$B impinging on $^{208}$Pb at 44 MeV/nucleon.
The ratio $\mathcal{R}_{\mathrm{sum}}$ and REB form factor $|F_{E,0}|^2$ (thick grey line) are considered at an energy $E=125$~keV in the $^7$Be-$p$ continuum and are given in units of MeV$^{-1}$. Differential breakup angular distributions $d\sigma_{\mathrm{BU}}/dEd\Omega$ are given in units of b/MeV~sr.
Calculations using different sets of potentials are displayed (see text for details).}
\end{figure}

We start our analysis of the ratio method at intermediate energies (44~MeV/nucleon) on a lead target; these correspond to the conditions of the MSU experiment of Davids \textsl{et al.} \cite{Dav98}.
On \Fig{f1} are represented the angular distributions for the breakup of $^8$B into $^7$Be and $p$ at the continuum energy $E=125$~keV (in b/MeV~sr), the summed cross section \eq{eq:Xsum} divided by Rutherford, their ratio $\mathcal{R}_{\mathrm{sum}}$ \eq{eq:ratio} expressed in MeV$^{-1}$, and the corresponding REB form factor $|F_{E,0}|^2$ [thick grey line, see \Eq{eq:formf}].
The solid curves correspond to the full calculation, including both $^7$Be-Pb and {$p$-Pb} interactions.
Calculations which do not include the latter are represented by the dashed red lines ($V_{pT}=0$) and calculations using the alternative $^7$Be-Pb potential ($V'_{cT}$, see \Sec{optpo}) are represented with the dash-dotted lines.

As initially observed in \Ref{capel2010} for a one-neutron halo projectile, the breakup and summed cross sections oscillate and decrease as functions of the scattering angle in a very similar pattern. Accordingly, their ratio removes most of these features, leading to a smooth curve that shows a similar trend as the REB prediction $|F_{E,0}|^2$ \cite{capel2011, capel2013}.
This result exhibits little dependence on the choice of the $^7$Be-Pb interaction: the ratio obtained with the alternative potential $V'_{cT}$ is nearly superimposed on the first one.
At forward angles, this simply reflects the fact that both potentials lead to indistinguishable cross sections, which is to be expected for a Coulomb-dominated reaction.
However, at angles $\theta\gtrsim8^\circ$, where the reaction becomes slightly more sensitive to the choice of nuclear potential and the individual cross sections exhibit noticeable differences, both ratios remain superimposed.
This result shows that the independence of the ratio to the optical-potential choice is also observed for loosely bound proton-rich nuclei.

Compared to the one-neutron halo cases studied in \Ref{capel2013}, the form factor predicted by the REB for $^8$B overestimates the DEA calculations (compare \Fig{f1} with the Figs.~2(b), 6 and 7 of \Ref{capel2013}).
To understand this difference, we test the two approximations included in the REB, meaning the effect of the $p$-$T$ interaction and the adiabaticity hypothesis.
When $V_{pT}$ is set to zero, the ratio superimposes nearly perfectly with the REB form factor. 
Additional tests---not plotted here for the sake of clarity---have shown that this difference is solely due to the Coulomb $p$-$T$ interaction.
We can explain this result by noting that, in the REB model, the breakup is caused by the sole recoil of the core due to its interaction with the target, the valence nucleon being seen as a spectator.
Unlike in one-neutron halo nuclei, the halo nucleon is charged here, which implies that the repulsive Coulomb interaction between this valence proton and the target reduces the tidal force, which is responsible for the dissociation.
The actual breakup and hence the ratio are then smaller than those predicted by the REB at forward angles.

As noted in previous work \cite{capel2011, capel2013}, the adiabatic approximation made in the REB is responsible for an additional overestimation of the actual ratio by the REB prediction (see, e.g., the inset of Fig.~2(b) and Fig.~6 of \Ref{capel2013}).
However, that adiabatic effect takes place only at very forward angles.
This is also observed here: at angles $\theta\lesssim0.5^\circ$, the REB form factor is slightly larger than the DEA ratio obtained with $V_{pT}=0$.
As expected from the very small binding energy of $^8$B, adiabaticity of the reaction assumed within the REB is rather well fulfilled here.

These results show that the overestimation of the ratio observed for $^8$B with a lead target is mainly due to the Coulomb repulsion that exists between the proton halo and the target.
This puts the ratio method at stake.
However, since the Coulomb interaction is significantly reduced with a light target, we investigate whether the aforementioned problem can be avoided on a carbon target.

\subsection{$^{12}$C target at 44 MeV/nucleon}\label{8BC}

\begin{figure}[htb]
	\centering
	\includegraphics[width=0.7\columnwidth]{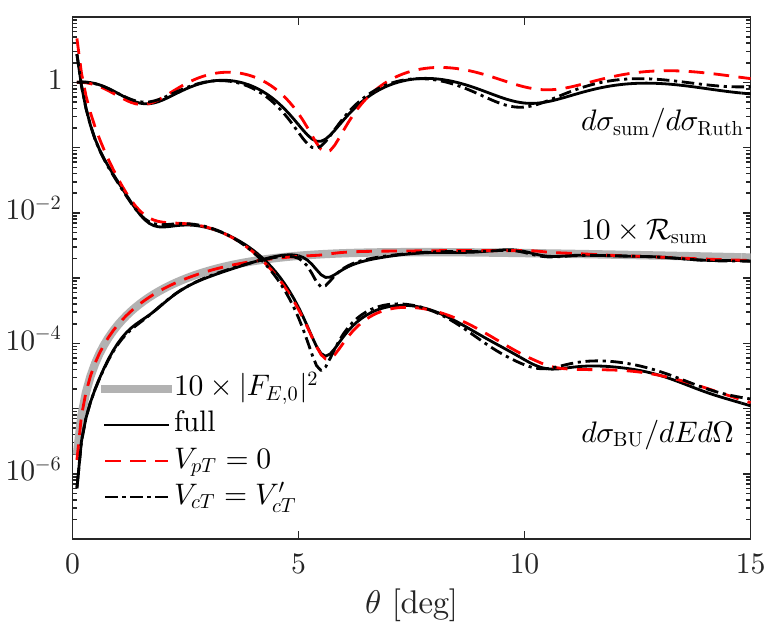}
	\caption{\label{f2} Same as Fig.~\ref{f1} but for a $^{12}$C target.
	Note that the ratio is multiplied here by 10 to improve the readability.}
\end{figure}

The results obtained on a C target at 44~MeV/nucleon are shown in \Fig{f2}.
As observed before, the breakup and summed cross sections oscillate in a very similar pattern.
By taking the ratio, these oscillations are strongly reduced.
However, contrarily to the Coulomb-dominated case, some remnant oscillations appear in the ratio.
This is very similar to what has been observed for neutron-halo nuclei \cite{capel2013}; the remnant oscillations are due to the slight shift that exists between the elastic and breakup angular distributions, which arises from the kick given by the target to the valence proton through $V_{pT}$ \cite{johnson1997,RCJ1997}.
This is confirmed by the calculation in which the $p$-$T$ interaction is neglected, which is in perfect agreement with the REB prediction.

At forward angle (viz. $\theta\lesssim3^\circ$), the REB form factor overestimates the full DEA calculation.
This is reminiscent of the problem observed with the Pb target.
Accordingly it has the same root, viz. the Coulomb \mbox{$p$-$T$} interaction, which hinders the breakup and leads to a larger REB prediction compared to the actual ratio.
However, the reactions on $^{12}$C being nuclear dominated, this reduction happens only at forward angles and is smaller than in a Coulomb-dominated reaction.

To study the independence of the ratio to the choice of the $c$-$T$ interaction, we use the alternative $^7$Be-$^{12}$C potential mentioned in \Sec{optpo} ($V'_{cT}$, dash-dotted lines). 
Unfortunately, both potentials provide nearly identical angular distributions.
Nevertheless, at large angles (i.e., for $\theta\gtrsim8^\circ$) they produce noticeable differences in the cross sections that are completely washed out within the ratio, confirming again that this observable removes most of the sensitivity to the $c$-$T$ optical potential.

\subsection{$^{58}$Ni target at 44 MeV/nucleon}\label{8BNi}
\begin{figure}[htb]
	\centering
	\includegraphics[width=0.7\columnwidth]{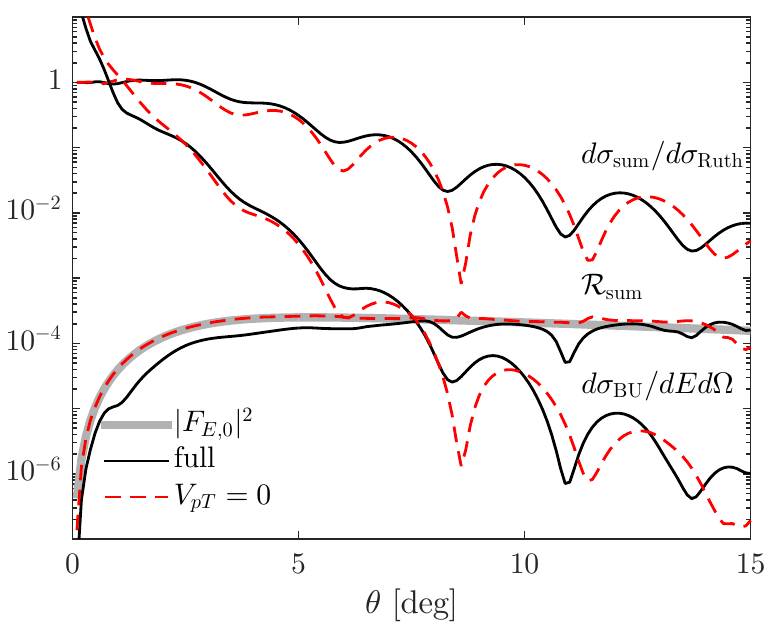}
	\caption{\label{f3} Same as Fig.~\ref{f1} but for a $^{58}$Ni target.}
\end{figure}

The previous sections have shown that the dynamical calculation can be directly confronted to the REB prediction only for a light target, like $^{12}$C.
Unfortunately, this is also the target that leads to the lowest breakup cross section and hence for which the ratio will be the hardest to measure.
In order to find a compromise between accuracy of the method and feasibility of the measurement, we have performed another series of calculations on a $^{58}$Ni target at the same energy.
The results are displayed in \Fig{f3}.

The results obtained on the $^{58}$Ni target are similar to those computed with the two previous targets.
Here also, the dynamical breakup and summed angular distributions exhibit very similar decays and oscillatory patterns, which roughly cancel out when considering their ratio.
However, that ratio is in less good agreement with the REB form factor than on $^{12}$C: at forward angle ($\theta\lesssim5^\circ$), the DEA ratio lies below its REB prediction, while at larger angles, it exhibits remnant oscillations.
Both problems fully disappear when the calculation is performed without the $p$-$T$ interaction.
As discussed in \Sec{8BPb}, the former issue is due to the dominance in that angular region of the Coulomb part of the $p$-$T$ interaction, which hinders the breakup.
The second issue is related to the whole $V_{pT}$, which produces a shift in the angular distributions, as explained in \Sec{8BC}.

Although it produces a larger breakup cross section than C for $\theta\lesssim 8^\circ$, which would thus be easier to measure, the Ni target does not seem the optimal choice for a measurement of the ratio for $^8$B because the ratio it produces cannot be directly related to the REB prediction.
At least for this nucleus, it seems that light targets should be favored in an experimental use of the ratio method.

\begin{figure}[htb]
	\centering
	\includegraphics[width=0.7\columnwidth]{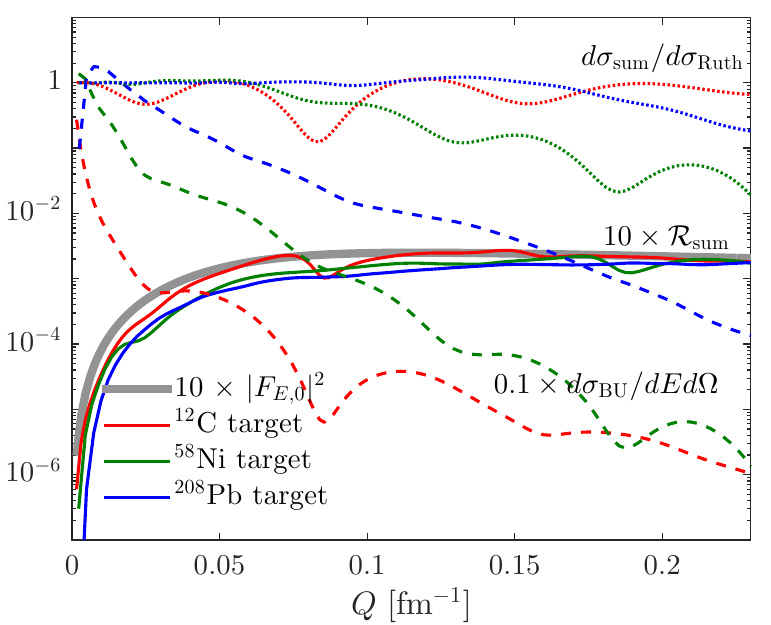}
	\caption{\label{f4} Sensitivity of the ratio to the target choice: DEA calculations on $^{12}$C, $^{58}$Ni, and $^{208}$Pb at 44~MeV/nucleon are displayed as a function of $Q$ [see \Eq{eq:Q}]. The breakup angular distributions, the ratio ${\mathcal R}_{\rm sum}$ and the REB form factor $|F_{E,0}|^2$ (thick grey line) are calculated at an energy $E=125$~keV in the $^7$Be-$p$ continuum.}
\end{figure}

To conclude this first series of tests of the ratio method extended to proton-halo nuclei, let us compare the ratios obtained with the three different targets with one another.
This is done in Fig.~\ref{f4}, where we have plotted the breakup angular distributions computed at $E=125$~keV expressed in b/MeV~sr (dashed lines) and the summed cross sections displayed as the ratio to Rutherford (dotted lines) together with their ratio (given in MeV$^{-1}$ but multiplied by 10 for readibility; solid lines) for the C (red lines), Ni (green lines), and Pb (blue lines) targets as a function of the momentum transfer $Q$ [see \Eq{eq:Q}].
We observe that even though the processes involved in the three collisions are very different and, accordingly, that the summed and breakup cross sections exhibit very different behaviours, all three ratios are pretty similar to one another. This confirms that, as observed for neutron-halo nuclei, the ratio method removes most of the dependence on the reaction mechanism. However, there remains a target dependence more significant than for neutron-halo nuclei \cite{capel2011, capel2013,colomer2016}.
Due to the presence of the $p$-$T$ Coulomb interaction, the experimental application of the ratio method in its original idea should be preferentially made on light targets.
On heavier targets, measurements should be compared to fully dynamical calculations, which, like the DEA, properly include $V_{pT}$, or to an extension of the REB, which is currently under development to include a perturbative estimate of $V_{pT}$ \cite{Johnson-private}.

\section{\label{sec:Sensitivity}Sensitivity of the ratio observable to the projectile structure}

\subsection{\label{sec:sensitivityDescription}Sensitivity to the $^7$Be-$p$ potential choice}

\begin{figure}[htb]
	\centering
	\includegraphics[width=0.7\columnwidth]{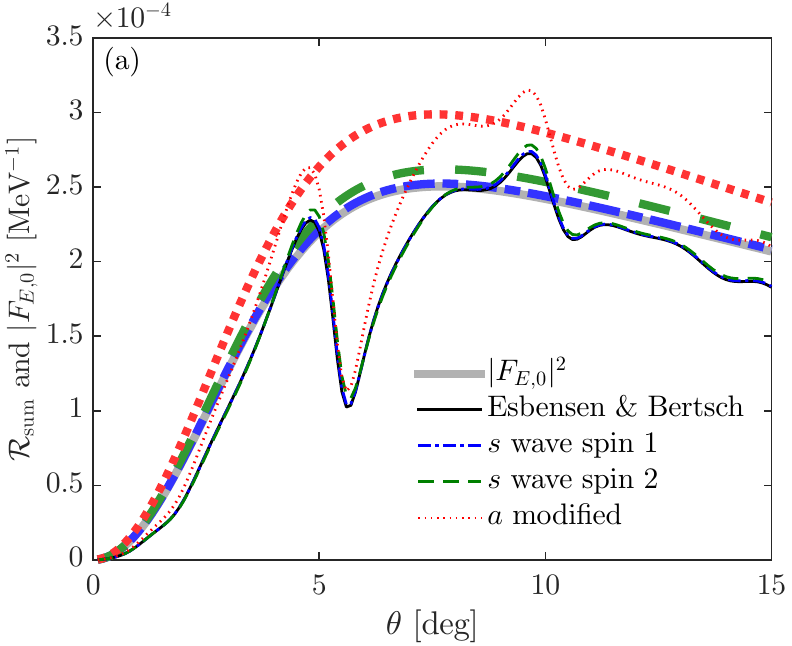}
	\caption{\label{ff5} 
		Sensitivity of ${\mathcal R}_{\rm sum}$ to the projectile description ($^8$B impinging on $^{12}$C at 44~MeV/nucleon). Besides the Esbensen and Bertsch $^8$B potential (black), potentials modified in the $s$ wave continuum that reproduce the scattering length of the spin-1 (blue) and spin-2 (green) channels, as well as a potential with a modified diffuseness (red) are also considered. Their corresponding form factor $|F_{E,0}|^2$ is given as thick lines of the same type and color.}
\end{figure}

To initiate our analysis of the sensitivity of the ratio ${\mathcal R}_{\rm sum}$ on the description of the projectile, we perform reaction calculations for $^{8}$B projectiles described by different $^{7}$Be-$p$ potentials all fitted to bind a $0p_{3/2}$ proton to the $^7$Be core by 137~keV.
We first test the sensitivity of the ratio to the low-energy $^{7}$Be-$p$ continuum using potentials in the $s$ wave that are adjusted to reproduce the scattering length in the spin 1 and 2 channels \cite{Barker}.
Second, we analyze the influence of the potential geometry upon the ratio by modifying the diffuseness of  the potential of Esbensen and Bertsch $a = 0.52$~fm to $a = 0.65$~fm.
Following the results from the previous section, we consider the most favorable case for the ratio method of a collision on $^{12}$C at 44~MeV/nucleon.
The corresponding DEA ratios (thin lines) alongside the REB form factors (thick lines) are displayed in \Fig{ff5}.

The results displayed in \Fig{ff5} in linear scale call for a general comment before discussing the details of the sensitivity of our calculations to the $^7$Be-$p$ potential.
We observe a less good agreement between the accurate DEA calculations and the REB prediction in this proton-rich case than for one-neutron halo nuclei (see, e.g., Fig.~4(b) of \Ref{capel2013}).
As mentioned in the previous section, at very forward angle (viz. $\theta\lesssim3^\circ$) the REB prediction does not precisely reproduce the dynamical calculations.
At larger angles, although the form factor predicts the general trend of the ratio, it misses the significant remnant oscillations observed in the DEA calculation. This is especially true for the dip observed at about 6$^\circ$.
As explained earlier, these oscillations arise from the proton-target interaction, which is neglected in the REB and more particularly from the shift between the summed and breakup cross section it causes.
Unlike for neutrons, this approximation is less valid in this charged case, leading to this less good agreement between the actual DEA calculation and its REB prediction.
To account for that difference between charged and neutral cases, we will suggest different ways to apply the ratio for proton-rich nuclei in \Sec{subsec:extension}.

In the analysis of the sensitivity of our calculations to the $^7$Be-$p$ potential, let us first observe that all ratios presented in \Fig{ff5} are very similar in shape and magnitude except for the potential with the larger diffuseness $a=0.65$~fm.
The influence of the continuum seems thus small compared to the one caused by the change in the potential geometry.
This is further confirmed if we set the $^7$Be-$p$ interaction to zero in all partial waves, but in the $0p_{3/2}$ (test not displayed here for clarity).

At the angles displayed here, the dependence to the potential geometry is mostly captured by the change in the asymptotic normalization coefficient (ANC) of the initial ground-state wave function.
Indeed, by dividing the ratios and REB form factors by the square of the ground-state ANC, all curves fall quite close to one another.
For a given binding energy, the ratio and REB form factor magnitude variations are thus mostly due to the differences in ANC.
This result is in good agreement with the results observed in the neutron-halo case \cite{capel2011,capel2013,colomer2016}.
The internal part of the projectile ground state wave function could be probed if the ratio was measured at sufficiently large scattering angle. However the ratio remains a mostly peripheral observable when measured at small angles, meaning that it probes the tail of the wave function, viz. its ANC.

\subsection{\label{sec:sensitivity}Sensitivity to the binding energy and orbital angular momentum of the valence proton}

\begin{figure}[htb]
	\centering
	\makebox[\textwidth][c]{\includegraphics[width=0.583\columnwidth]{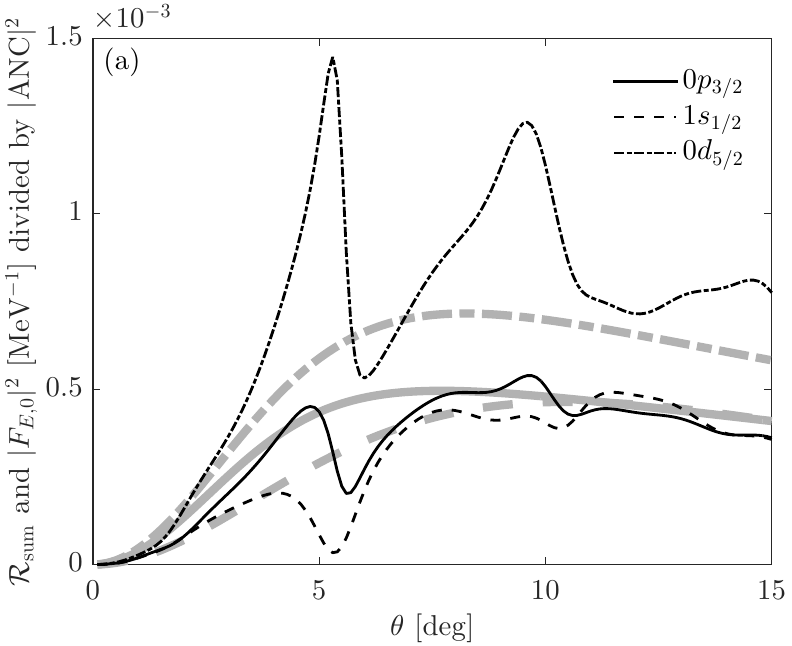}%
	\includegraphics[width=0.6\columnwidth]{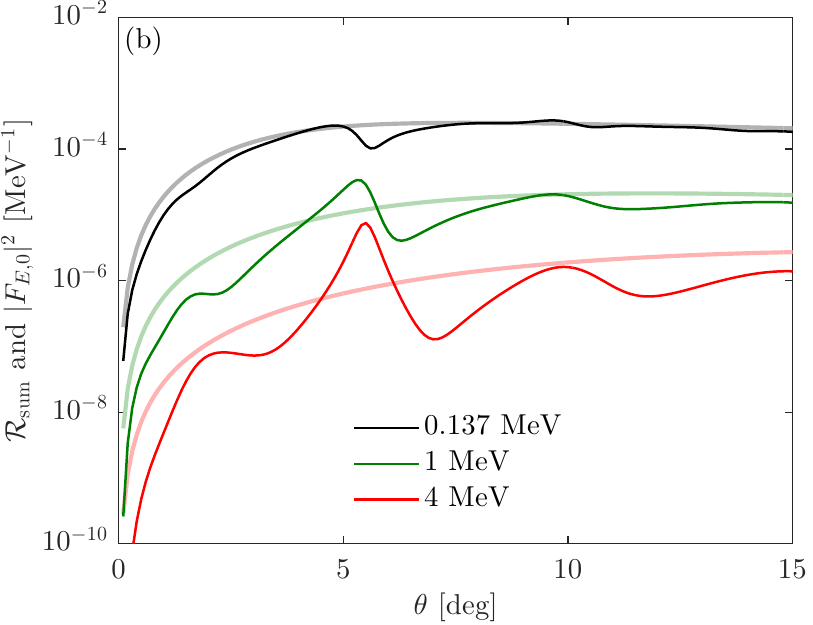}}
	\caption{\label{f5} 
	Sensitivity of ${\mathcal R}_{\rm sum}$ to the projectile structure ($^{12}$C target and 44~MeV/nucleon). Besides the realistic $^8$B (valence $p$ bound by 137~keV in the $0p_{3/2}$ orbit), projectiles with (a) different orbitals ($1s_{1/2}$ and $0d_{5/2}$), and (b) different ground-state energies ($|E_{0p3/2}|=1$ and 4~MeV) are also considered. Their corresponding form factor $|F_{E,0}|^2$ is given as thick lines of the same type and color.}
\end{figure}

Let us now evaluate the sensitivity of the ratio ${\mathcal R}_{\rm sum}$ to the orbital and binding energy of the valence proton.
To this aim, we follow Refs.~\cite{capel2011,capel2013,colomer2016} and perform calculations for $^8$B-like projectiles in which the valence proton is bound to the $^7$Be core within different orbitals and with different binding energies.
Again, we consider a collision on $^{12}$C at 44~MeV/nucleon.
The corresponding DEA ratios (thin lines) alongside the REB form factors (thick lines) are displayed in \Fig{f5}.
In addition to the physical $^8$B (a $0p_{3/2}$ proton bound by 137~keV to the $^7$Be core; solid black and grey lines), we consider $1s_{1/2}$ [dashed lines in \Fig{f5}(a)] and $0d_{5/2}$ [dash-dotted lines in \Fig{f5}(a)] valence protons bound by 137~keV as well as $0p_{3/2}$ states with binding energies of 1~MeV [green lines in \Fig{f5}(b)] and 4~MeV [red lines in \Fig{f5}(b)].
Since the ANC dominates the magnitude of the ratio for a given binding energy, we divide the ratios and form factors by the squared ANC of their respective ground-state wave function in \Fig{f5}(a), viz. by 5.81~fm$^{-1}$, 0.504~fm$^{-1}$, and 0.010~fm$^{-1}$, for the $1s_{1/2}$, $0p_{3/2}$, and $0d_{5/2}$ states, respectively.

As expected from Eq.~(\ref{eq:formf}), the form factor shows a significant dependence on the projectile structure.
As for neutron-halo nuclei \cite{capel2013,colomer2016}, this dependence is visible in the form-factor magnitude as well as in its shape.
Note that the square of the ANC decreases by one order of magnitude each time the ground-state orbital angular momentum $l_0$ is increased by one unit, which means that in addition to the change in shape seen in \Fig{f5}(a), a variation in orbital angular moment leads to an even larger change in the magnitude of the ratio.
The binding energy has a similar influence on the magnitude of the ratio.
Although the quality of the REB prediction is lower in the present case than for one-neutron halo nuclei, the significant changes illustrated in \Fig{f5} will enable experimentalists to infer 
pertinent structure information from the analysis of actual data.
Except for the remnant oscillations, the REB prediction for a projectile bound in the $s$ or $p$ wave follows fairly well the trend of the DEA ratio [see \Fig{f5}(a)]. For a projectile bound in the $d$ wave however, the agreement is not good. This confirms the results already observed for neutron halos \cite{capel2013,colomer2016}: the agreement between the REB form factor and the actual ratio is better for low orbital angular momentum $l_0$.
This agreement deteriorates when the binding energy increases [see \Fig{f5}(b)], indicating that the ratio method works at best for loosely bound systems, like halo nuclei.

This series of tests shows that although it works less well than for one-neutron halo nuclei, the ratio method can provide valuable information about the single-particle structure of unstable proton-rich nuclei, even if the direct comparison of data to the REB prediction \eq{eq:REBratio} will be less precise than for one-neutron halo nuclei.
In particular, when the binding energy or the orbital angular momentum increase, this direct confrontation will no longer be reliable and the data will thus have to be analyzed with an accurate model of the reaction, like the DEA (see \Sec{subsec:extension}).
 

\subsection{\label{sec:continuum}Choice of the continuum energy}

The calculations presented in the previous sections consider a single continuum energy $E=125$~keV between the $^7$Be core and the halo proton in the breakup channel.
In the present section, we study the influence of the choice of this energy on the ratio method.
In particular, we check how the method works at higher energy $E$ in the $c$-$p$ continuum and when an energy range is considered instead of a single energy.
We also study if the presence of a resonance in the continuum affects the method.

An actual measurement of the ratio will require to consider a continuum-energy range or \textit{bin}, and the statistics uncertainty will be improved if a broad bin can be considered.
We therefore analyze how the method is affected by such a binning and how it varies with the width of the energy range.
Namely, we consider the following bin ratio
\beq
\mathcal{R}_{\mathrm{sum}}(\mathrm{bin},\bm{Q}) &=& \int_{E_{\rm min}}^{E_{\rm max}} \mathcal{R}_{\mathrm{sum}}(E,\bm{Q})\ dE \label{eq:ratioint}\\
 &=&\frac{\int_{E_{\rm min}}^{E_{\rm max}}(d\sigma/dEd\Omega)_{\mathrm{bu}}\ dE}{(d\sigma/d\Omega)_{\mathrm{sum}}},
\eeqn{e11}
where $E_{\rm min}$ and $E_{\rm max}$ are respectively the lower and higher bounds of the bin.
This ratio is associated to the bin-integrated REB form factor
\beq
|F_{\mathrm{bin},0}(\bm{Q})|^2 &=&\int_{E_{\rm min}}^{E_{\rm max}} |F_{E,0}(\bm{Q})|^2\ dE.\eeqn{eq:formfint}

\begin{figure}[htb]
	\centering
	\makebox[\textwidth][c]{\includegraphics[width=0.57\columnwidth]{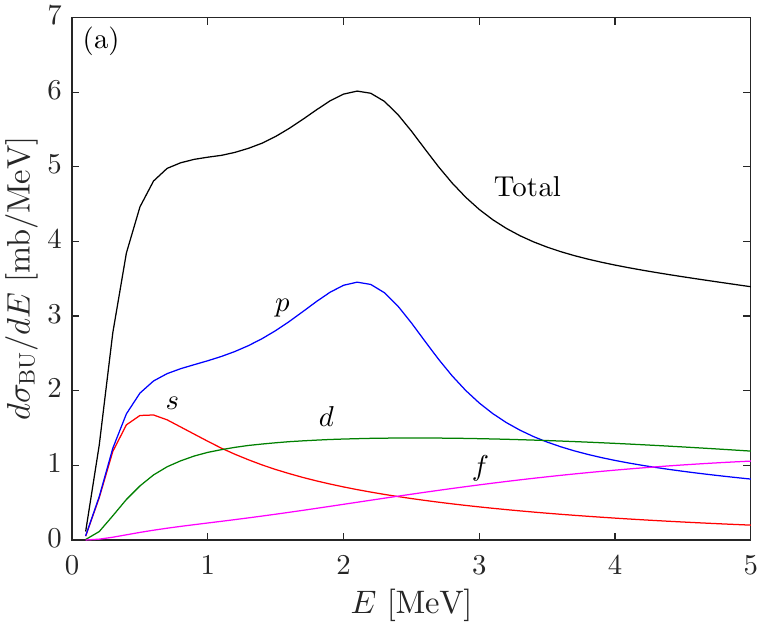}%
	\includegraphics[width=0.6\columnwidth]{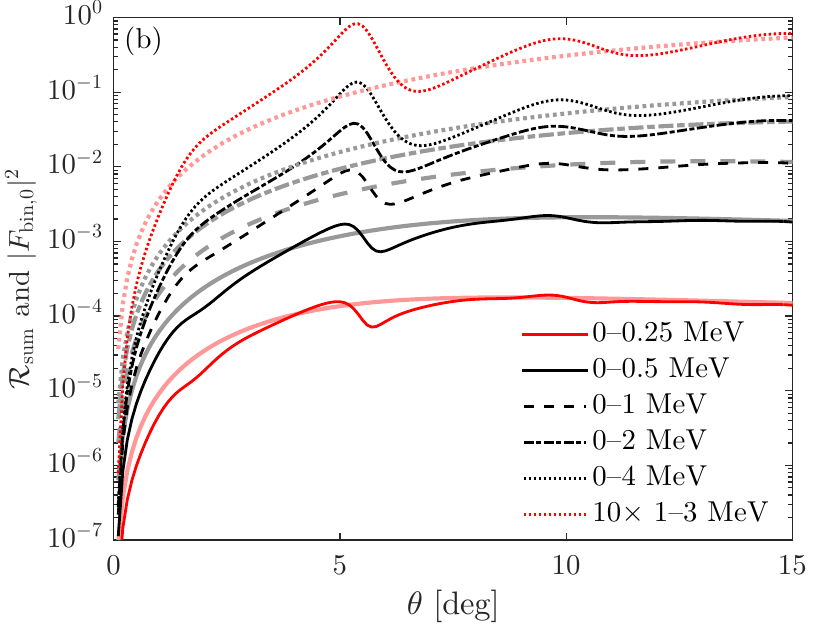}}
	\caption{\label{f6} 
	Choice of the continuum energy bin for the experimental exploitation of the ratio method.
(a) The breakup cross section plotted as a function of the continuum energy $E$.
(b) ${\mathcal R}_{\rm sum}$ computed within the DEA (thin lines) and its REB estimate (thick lines) for different energy bins in the $^{7}$Be-$p$ continuum.}
\end{figure}

The DEA energy distribution for the breakup of $^8$B on $^{12}$C at 44~MeV/nucleon is presented in \Fig{f6}(a), the contributions of the $s$, $p$, $d$, and $f$ partial waves in the continuum are shown separately.
The shape of this cross section is typical of the nuclear-dominated breakup reaction of halo nuclei \cite{CGB04,Fuk04}.

Opportunely, the simplified version of the $^7$Be-$p$ potential of Esbensen and Bertsch \cite{esbensen1996} presented in \Sec{8B} leads to a $p_{1/2}$ resonance at 2.3~MeV above the one-proton threshold with a width of 1.6 MeV.
Albeit unphysical, that state will enable us to study the behavior of the ratio on and off resonance.
As in Refs.~\cite{CGB04,Fuk04}, we observe a peak in the breakup cross section due to its $p_{1/2}$ contribution at the energy and with a width similar to those of the resonance.

For this study, we divide the $^7$Be-$p$ continuum into six different bins.
The first one, 0--0.25~MeV, is chosen at low energy and centered on the $E=125$~keV used in the previous sections.
The next two, 0--0.5~MeV and 0--1~MeV, are kept in the low-energy part of the non-resonant continuum.
The fourth bin, 0--2~MeV, includes part of the resonant range.
To clearly see the influence of the resonance, we also consider a bin centered on it: 1--3~MeV.
Finally, to test the possible use of a much broader bin, we look at the 0--4~MeV range.
The corresponding results are displayed in \Fig{f6}(b).

The first bin (0--0.25~MeV) provides a ratio very similar to that of the sole $E=125$~keV.
The ratio calculated with the second (0--0.5~MeV) bin is qualitatively identical to those two with the only quantitative difference that this larger bin leads to a breakup cross section---and hence a ratio---ten times as large as the single continuum energy, which would be useful in an experimental application of the method.

The third (0--1~MeV), and fourth (0--2~MeV) bins lead to similar results, although the disagreement between the DEA ratio and the REB form factor slightly increases with $E_{\rm max}$.
This suggests that the resonance does not really affect the ratio method.
This is confirmed by the calculation performed with the bin centered on the resonance (1--3~MeV), which does not exhibit any peculiar behavior compared to the non-resonant continuum.
The much broader bin (0--4~MeV) provides similar results.
For the last two bins, the form factor is in less good agreement with the DEA ratio.
This is expected because higher excitation energies are less compatible with the adiabatic approximation of the REB.
However, they provide a larger ratio, which could significantly improve the statistics uncertainty in actual data taking.
For practical purposes, a balance will thus have to be sought between the accuracy of the method and the practicality of its experimental implementation.

\subsection{Extension of the ratio to charged systems}\label{subsec:extension}
The results obtained so far show that the ratio method applied to charged systems is not as convincing as for one-neutron halo nuclei.
There remains a sensitivity to the target choice (see \Fig{f4}) and even on light targets, the ratio obtained in a fully dynamical reaction model exhibits remnant oscillations not observed within the REB prediction.
Nevertheless, as shown in \Sec{sec:8B}, the breakup and summed angular distributions exhibit very similar features, like Coulomb rainbow and oscillatory pattern.
By taking the ratio of the cross sections these features are mostly removed.
In addition, the ratio is completely independent of the optical potentials used to simulate the interaction between the core of the projectile and the target.
This in itself is interesting because the individual reaction observables are strongly sensitive to this choice and hence hinder the extraction of spectroscopic information from cross sections (see, e.g., \Ref{CGB04}).
Moreover as clearly illustrated in \Sec{sec:sensitivity}, the ratio varies enormously in both shape and magnitude with the core-halo orbital angular momentum and binding energy in the ground state.
Finally, as mentioned in \Ref{capel2011}, being the ratio of two cross sections, this particular observable is not sensitive to their normalization, which would be an appreciable quality in its experimental measurement.

It seems therefore that albeit less accurate than for neutron halos, the ratio exhibits enough advantages to make it useful to infer information on the proton-rich side of the valley of stability.
As seen in the previous sections, this is particularly true when the valence proton is loosely bound to a core in a low orbital angular momentum, viz. for proton halos.
We see here three variants of the method that can be applied in practice to benefit from this new observable.

In its \emph{strict application}, i.e. the one suggested in Refs.~\cite{capel2011,capel2013,colomer2016} for the study of one-neutron halo nuclei, an experimental measurement of the ratio is to be compared directly to the REB form factor \eq{eq:REBratio}.
As observed in Secs.~\ref{sec:sensitivityDescription} and \ref{sec:sensitivity}, even on a light target, there is too large a difference between the DEA ratio and its REB prediction to allow for this comparison to provide us with fine details about the structure of the projectile, at least for $^8$B.

Nevertheless, since the ratio varies by orders of magnitude with the projectile structure (see \Fig{f5}), pertinent information pertaining to that structure could already be obtained by confronting the order of magnitude and general shape of the experimental ratio to the REB prediction.
This \emph{approximate application} of the ratio would not be able to distinguish small differences in the ANC like the one illustrated in \Fig{ff5}, but it could still provide a good estimate of $l_0$ and $E_0$, which are difficult to measure directly far from stability.

In a third application of the ratio, which we coin \emph{dynamical}, measurements are compared not to the REB prediction, but to the results of state-of-the-art dynamical reaction calculations.
The gain in this case over the more usual analysis of cross sections for individual reactions lies in the complete independence of the ratio to the choice of $V_{cT}$.
This is of course the least practical use of the ratio since it requires an accurate calculation of the reaction, which is cumbersome to perform.
However, it is the one that would provide the most precise information about the single-particle structure of proton-rich nuclei far from stability.

Note also that since the major part of the disagreement between the REB form factor and the actual DEA ratio comes from the Coulomb part of the $p$-$T$ interaction, accounting for that interaction, e.g., at the first order of the perturbations, would improve the REB prediction and enable the direct comparison suggested in Refs.~\cite{capel2011,capel2013,colomer2016}.

In the next section, we explore how the ratio behaves when applied on other proton-rich nuclei with a clear single-particle structure.
In each case, we investigate which of these different applications of the ratio method can be used in practice.

\section{Extension of the ratio method to other proton-rich nuclei}\label{sec:otherhalo}

\subsection{\ex{17}{F}, \ex{25}{Al}, and \ex{27}{P}}\label{subsec:otherhalo}

After the detailed examination of the \ex{8}{B} case, we check the applicability of the ratio method to other proton-rich nuclei. The cases studied are \ex{17}{F}, \ex{25}{Al}, and \ex{27}{P}. These proton-rich $s$-$d$ nuclei are all seen as composed of a core of spin nil ($^{16}$O, $^{24}$Mg, and $^{26}$Si, respectively) and a loosely bound valence proton.

The loosely bound nucleus $^{17}$F ($S_p=601$~keV) exhibits a $\frac{5}{2}^+$ ground state.
In addition it also has a $\frac{1}{2}^+$ bound excited state at 106~keV below the one-proton separation threshold, which is usually depicted as exhibiting a one-proton halo structure.
Within an extreme shell model, they are seen as a $0d_{5/2}$ and $1s_{1/2}$ proton bound to an $^{16}$O core. This vision has recently been confirmed by a coupled-cluster calculation by Hagen \etal\ \cite{Hag10}.
In addition to these two bound states, $^{17}$F exhibits a $\frac{3}{2}^+$ resonance at 4.4~MeV above the one-proton threshold.
It is seen as the $d_{3/2}$ spin-orbit partner of the ground state.
Sparenberg, Baye and Imanishi have developed an \mbox{$^{16}$O-nucleon} potential that describes the low-energy spectra of the mirror nuclei $^{17}$F and $^{17}$O \cite{SBI00}.
This potential includes a central plus a spin-orbit terms of Woods-Saxon form factor, and reproduces the three aforementioned states.

Being closer to the proton dripline, the other nuclei have a less-known structure.
The one-proton separation energy of $^{25}$Al is 2.272~MeV and its ground state has spin and parity $\frac{5}{2}^+$. It is therefore seen as a proton bound to a $^{24}$Mg core in the $0d_{5/2}$ orbit. 
Interestingly, $^{27}$P exhibits a $\frac{1}{2}^+$ ground state seen as a $1s_{1/2}$ proton bound by 870~keV to a $^{26}$Si core.
Following the results of \Sec{sec:Sensitivity}, thanks to its loose binding and the nil orbital angular momentum of its valence proton, this nucleus could be an interesting test case for the ratio method close to the proton dripline.

To describe these nuclei within the two-cluster model presented in \Sec{model} and to study the influence of that choice of description upon the ratio method applied to these nuclei, we consider, for each one of them, two sets of $c$-$p$ potential, $V_{cp}$ and $V'_{cp}$.
For $^{17}$F, we use the potential developed by Sparenberg \etal\ in \Ref{SBI00}, whose parameters are given in the first line of Table~\ref{tab-1}.
As second potential $V'_{cp}$, we use the Woods-Saxon geometry of the $^7$Be-$p$ potential of Esbensen and Bertsch \cite{esbensen1996}---viz. with $a=0.52$~fm and $r_0=1.25$~fm---adjusting the central depth to bind the valence proton at the right energy in the $0d_{5/2}$ orbit.
For simplicity, we ignore the spin-orbit splitting.
The parameters of that potential are listed in the second line of \tbl{tab-1}.
In that potential, the $1s_{1/2}$ is bound by 760~keV, hence below the ground state, and the $0d_{3/2}$  state is degenerated with the $0d_{5/2}$ state.

\begin{table}
	\setlength{\tabcolsep}{5pt}
	\caption{\label{tab-1} Parameters of the single-particle potentials $V_{cp}$ and $V'_{cp}$ used to describe $^{17}$F, $^{25}$Al, and $^{27}$P. The corresponding orbital and experimental binding energy of the valence proton in the ground state of these nuclei are also listed \cite{Firestone-NDS-2009, Basunia-NDS-2011}.}
	\begin{indented}
	\item[]\begin{tabular}{c c c c c c c}\hline\hline
	& $V_{0}$  &$V_{\rm LS}$  & $a$   &$r_{0}$ & $n_0l_0j_0$  &$E_0$  \\
	& (MeV)     & (MeV fm$^2$)& (fm)   & (fm)     &                       & (MeV) \\\hline
	\multirow{2}{*}{\nuc{17}{F}}  & 56.700     &25.14  & 0.642 &1.20   &\multirow{2}{*}{$0d_{5/2}$} & \multirow{2}{*}{0.600} \\
	                     & 57.090           &0     & 0.52          &1.25   & & \\\hline
	\multirow{2}{*}{\nuc{25}{Al}}  & 50.342           &0     & 0.65          &1.25   &\multirow{2}{*}{$0d_{5/2}$} & \multirow{2}{*}{2.272} \\
	                     & 49.346           &0     & 0.52          &1.25   & & \\\hline
	\multirow{2}{*}{\nuc{27}{P}}   & 47.377           &0     & 0.65          &1.25   &\multirow{2}{*}{$1s_{1/2}$} & \multirow{2}{*}{0.870} \\
	                     & 48.136           &0     & 0.52          &1.25   & & \\\hline\hline
	\end{tabular}
	\end{indented}
\end{table}

To describe $^{25}$Al and $^{27}$P, we consider simple Woods-Saxon potentials without spin-orbit term.
The first $V_{cp}$ is chosen with the usual diffuseness $a=0.65$~fm and reduced radius $r_0=1.25$~fm.
For each nucleus, its depth is adjusted to reproduce the experimental one-proton separation energy in the physical partial wave (see lines 3 and 5 of \tbl{tab-1}); the same potential is considered in all partial waves.
To get a second $c$-$p$ potential $V'_{cp}$, we do as for $^{17}$F and consider the geometry of the potential of Esbensen and Bertsch \cite{esbensen1996}, adjusting its depth to reproduce the correct binding energy (see lines 4 and 6 of \tbl{tab-1}).

To study the application of the ratio method to these nuclei, we follow the results of \Sec{sec:8B} and first consider a carbon target.
We then analyze how choosing a Ni target affects the method.
The beam energy is selected at 60~MeV/nucleon, which can be produced at various RIB facilities. The optical potentials used in these tests are the systematic nucleus-nucleus potential of Xu and Pang \cite{XP13} for $V_{cT}$ and the Chappel-Hill global nucleon potential for $V_{pT}$ \cite{CH89}.
The former is obtained by folding the effective JLMB nucleon-nucleon interaction \cite{Bauge-PRC-2001} with the nucleon density distributions of the projectile and target obtained with Hartree-Fock calculations using the SkX interaction \cite{Brown-PRC-1998}.
For all three projectiles, the real and imaginary parts of the nucleus-nucleus potential are renormalized with the factors $N_r=0.68$ and $N_i=1.22$, respectively.
To study the effect of different $c$-$T$ interactions on the ratio method, calculations are also made with another $V_{cT}'$, which is arbitrarily chosen to have $N_r=0.58$ and $N_i=1.02$. 

\subsection{\label{FAlPC}$^{12}$C target at 60~MeV/nucleon}

\begin{figure*}[htbp]
	\centering
	\makebox[\textwidth][c]{\includegraphics[width=1.35\textwidth]{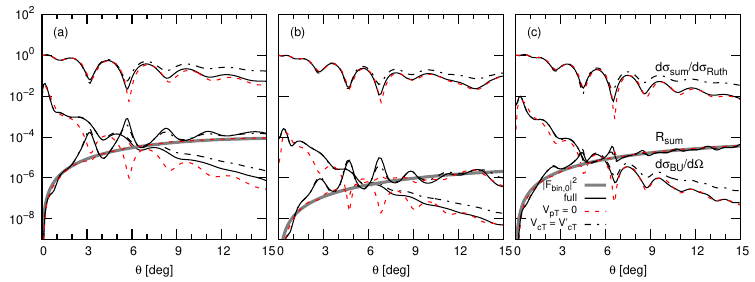}}
	\caption{Analysis of the ratio method for (a) \nuc{17}{F}, (b) \nuc{25}{Al}, and (c) \nuc{27}{P} impinging on a \nuc{12}{C} target at 60 MeV/nucleon.
		The ratio $\mathcal{R}_{\mathrm{sum}}$ and REB form factor $|F_{E,0}|^2$ (thick grey line) are considered at the bin energy of $E= 0.0$--1.0 MeV in the $c$-$p$ continuum and have no units. Differential breakup angular distributions over that bin $d\sigma_{\mathrm{BU}}/d\Omega$ are given in units of b/sr.
		Calculations using different sets of potentials are displayed (see text for details).}
	\label{f7}
\end{figure*}

\begin{figure*}[htbp]
	\centering
	\makebox[\textwidth][c]{\includegraphics[width=1.35\textwidth]{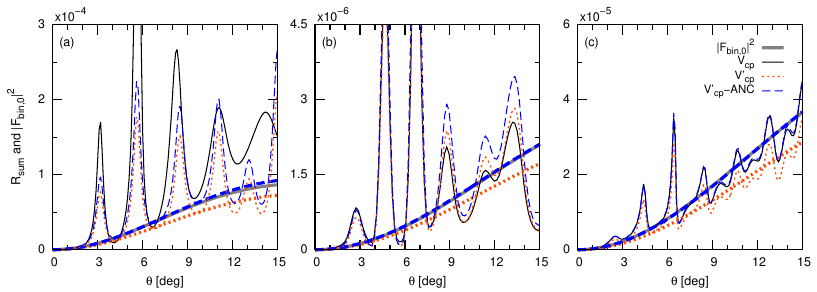}}
	\caption{Sensitivity of $\mathcal{R}_\mathrm{sum}$ to the projectile description for (a)~\nuc{17}{F}, (b)~\nuc{25}{Al}, and (c)~\nuc{27}{P} impinging on \nuc{12}{C} at 60~MeV/nucleon. The ratio $\mathcal{R}_\mathrm{sum}$ (thin lines) and REB form factor $|F_{\mathrm{bin},0}|^2$ (thick lines) are considered at the bin energy of $E= 0$--1 MeV in the $c$-$p$ continuum.}
	\label{f8}
\end{figure*}

Following what has been done in Secs.~\ref{8BPb}--\ref{8BNi}, we first analyze how the ratio method works for each of the nuclei considered in this study.
Figure~\ref{f7} displays our results for the collision of (a)~$^{17}$F, (b)~$^{25}$Al, and (c)~$^{27}$P on a $^{12}$C target at 60 MeV/nucleon; the first potential $V_{cp}$ is used in this study, the influence of that choice on our calculations is analyzed later in \Fig{f8}.
As before, the summed cross section \eq{eq:Xsum} is plotted as a ratio to Rutherford.
Following the results of \Sec{sec:continuum}, the breakup cross section is depicted for a 0--1~MeV bin in the $c$-$p$ continuum (expressed in b/sr) and their ratio $\mathcal{R}_\mathrm{sum}$ \eq{eq:ratio} is compared to the REB form factor $|F_{\mathrm{bin},0}|^2$ \eq{eq:REBratio} (thick grey line).
We represent the full calculations including both the $c$-$T$ and $p$-$T$ interactions (solid lines), those that do not include the latter ($V_{pT}=0$, dashed red lines), and those using the second $c$-$T$ potential ($V'_{cT}$, dash-dotted lines).

Contrary to what has been seen for $^{8}$B in \Sec{sec:8B} and for one-neutron halo nuclei in \Ref{capel2010}, we observe that for $^{17}$F and $^{25}$Al, the breakup and summed cross sections are totally out of phase.
Consequently, their ratios exhibit very strong oscillations, indicating that for these nuclei the method does not fully remove the dependence on the reaction process.
For $^{27}$P, on the contrary, we observe that both cross sections follow a very similar pattern and that the DEA ratio follows more closely its REB prediction than $^{17}$F and $^{25}$Al.
This confirms the results of \Sec{sec:sensitivity}, where we have seen that the method works best for valence protons loosely bound in an $s$ or $p$ orbital.

When the $p$-$T$ interaction is neglected, $\mathcal{R}_\mathrm{sum}$ falls very close to $|F_{\mathrm{bin},0}|^2$.
This is especially true for $^{27}$P and $^{17}$F.
For $^{25}$Al, there remain significant oscillations.
We interpret this as due to the fact that its valence proton is more deeply bound and sits in a $d$ wave, which is the less favorable case to apply the ratio method.

Interestingly though, the ratios obtained in all three cases exhibit little dependence on the $c$-$T$ optical potential.
When the reaction calculations are performed with the alternate potential $V'_{cT}$, although significant differences are seen in the individual cross sections, their ratios remain unchanged.
Once again, this is especially true for $^{27}$P for which the method works best.

These results confirm both the interest and limitation of the ratio method applied to proton-rich nuclei.
In the vocabulary developed in \Sec{subsec:extension}, the \emph{approximate} or even the \emph{strict application} of the ratio could be used for $^{27}$P.
For the $d$ bound states, on the contrary, there remain significant effects of the reaction dynamics and the ratio could at best be applied in its \emph{approximate} version.
Note however that, even for $^{25}$Al, which is more deeply bound in the $d$ wave, some information might be gathered by focussing on the order of magnitude and general shape of the ratio.


In \Fig{f8}, we study the influence on the ratio method of the $c$-$p$ potential used to describe (a)~$^{17}$F, (b)~$^{25}$Al, and (c)~$^{27}$P.
The linear scale enables us to focus on the details of these calculations.
For the $d$-bound nuclei, $^{17}$F and $^{25}$Al, the REB form factor merely reproduces the order of magnitude of the DEA ratio.
Hence, at best an estimate of $l_0$ and/or $E_0$ could be inferred from an \emph{approximate} version of the ratio.
More detailed information might be obtained using its \emph{dynamical} version.
On the contrary, the ratio method applied to $^{27}$P leads to a fair agreement between the DEA calculation and its REB prediction.
Although it does not exhibit the perfectly smooth angular dependence of the latter, the DEA ratio merely oscillates around the REB, which suggests that in this case the \emph{strict application} of the ratio would be possible.
This is especially true if the experimental angular resolution is not too fine.

In addition to the calculations performed with the first $V_{cp}$ potentials (solid black and grey lines), we also display the ratios obtained with $V'_{cp}$ (red dotted lines).
To understand to what the differences between both sets of calculations are due, we plot with blue dashed lines the $V'_{cp}$ ratios and form factors normalized to the $V_{cp}$ ANC---viz. we multiply them by $|\mathrm{ANC}_{V_{cp}}|^2/|\mathrm{ANC}_{V'_{cp}}|^2$.
For $^{27}$P, the scaled results are superimposed on the original $V_{cp}$ calculations, confirming that the reaction process is peripheral and probes only the tail of the ground-state wave function.
Moreover, since the REB prediction follows so closely the DEA ratio, the method could be used in its \emph{strict} version.
Were the experimental uncertainty sufficiently low, the ANC could probably be inferred by confronting the data to the REB form factor.

On the contrary, for $^{17}$F and $^{25}$Al, the choice of the interaction affects the quality of the method.
Although the scaled REB form factors are on top of the $V_{cp}$ ones, this is not the case for the DEA ratios.
The reaction process for these projectiles is consequently less peripheral, meaning that it is sensitive to the internal part of the wave function, which is not unexpected for systems bound in a $d$ wave.
Since this is not observed in the form factor, we know that dynamical effects spoil the original idea of the ratio method \cite{capel2011}, which is based on the REB that makes the adiabatic approximation.
This explains why, for these nuclei, the ratio method works only at the qualitative level and that only its \emph{approximate} version could be used.
Fine details on their structure, like the ANC of their ground state cannot be inferred with this method.

\begin{figure*}[htbp]
	\centering
	\makebox[\textwidth][c]{\includegraphics[width=1.35\textwidth]{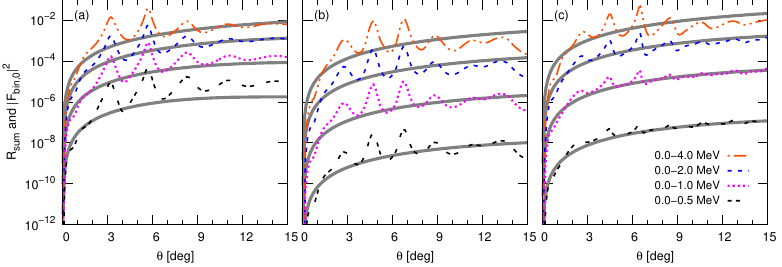}}
	\caption{Same as Fig.~\ref{f6}(b) but for (a) \nuc{17}{F}, (b) \nuc{25}{Al}, and (c) \nuc{27}{P} impinging on a \nuc{12}{C} target at 60~MeV/nucleon.}
	\label{f9}
\end{figure*}

To complete this study, we analyze in \Fig{f9} the best choice of continuum bin upon which to apply the ratio method.
These plots confirm that this method applied to $^{17}$F [\Fig{f9}(a)] and $^{25}$Al [\Fig{f9}(b)] is not accurate and provides about the same agreement between the DEA calculation and the REB form factor at all energies in the continuum.
Nevertheless, since the order of magnitude is well reproduced, the method could be used in its \emph{approximate} version.
For $^{27}$P, however, we observe as in \Fig{f6}(b), that the agreement between the DEA ratio and its REB estimate deteriorates when the continuum energy increases, which is what is expected in a rigorous application of the method \cite{capel2013,colomer2016}.
Since the breakup cross section increases significantly with the size of the energy bin, it will be necessary, in an experimental use of the method, to consider it broad enough.
As for $^8$B, the range $E=0$--1~MeV seems optimal having a DEA ratio quite close to the form factor while providing a breakup cross section two orders of magnitude larger than the $E=0$--0.5~MeV bin.

\begin{figure*}[htbp]
	\centering
        \includegraphics[width=0.5\columnwidth]{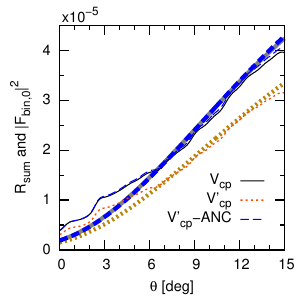}
	\caption{Effect of a coarse experimental angular resolution on the ratio applied to \nuc{27}{P}. The DEA ratios and their REB prediction shown in \Fig{f8}(c) have been averaged over a $4^\circ$ range to simulate a poor experimental resolution.}
	\label{f8b}
\end{figure*}

In order to test the aforementioned idea of measuring the ratio with a coarse angular resolution, we display in \Fig{f8b} the calculations on $^{27}$P shown in \Fig{f8}(c) averaged over a $4^\circ$ range to simulate a coarse angular resolution.
The REB prediction is now in very good agreement with the DEA ratio, confirming that in these conditions---light target, broad energy bin in the continuum and coarse angular resolution---the ratio method could be used in its \emph{strict} version and that, for this nucleus, fine details of its structure may be obtained from experiment.

\subsection{\label{FAlPNi}$^{58}$Ni target at 60~MeV/nucleon}

\begin{figure*}[htbp]
	\centering
	\makebox[\textwidth][c]{\includegraphics[width=1.35\textwidth]{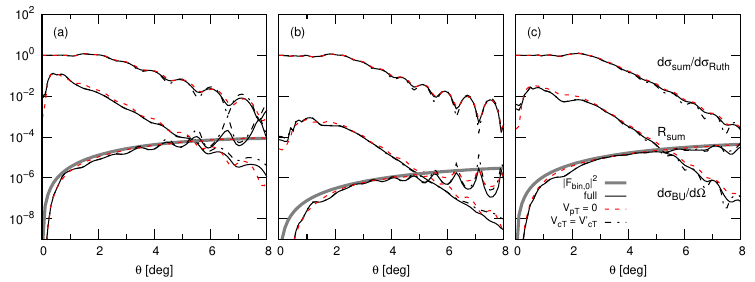}}
	\caption{Same as Fig.~\ref{f7} but on a \nuc{58}{Ni} target at 60~MeV/nucleon.}
	\label{f10}
\end{figure*}

To see if another target choice could increase the magnitude of the breakup cross sections and hence ease the experimental use of the ratio method, we perform the same calculations considering a $^{58}$Ni target at the same 60~MeV/nucleon beam energy.
The corresponding summed and breakup cross sections are displayed in Fig.~\ref{f10} as well as the DEA ratio and its REB prediction (thick grey line); the calculations correspond to the first $V_{cp}$ listed in \tbl{tab-1}.

As expected, the larger Coulomb $P$-$T$ interaction leads to a larger breakup cross section.
It also significantly affects the general shape of the angular distributions, both the summed and breakup ones.
We see the clear appearance of a Coulomb rainbow and the shift of the oscillatory pattern to larger angles.
However, as already observed on the $^{12}$C target, for $^{17}$F and $^{25}$Al the summed and breakup cross sections do not exhibit the same behavior showing that for these nuclei, the ratio method can be qualitative at best.
For $^{27}$P however, both cross section behave similarly, leading to a rather smooth DEA ratio.
Unfortunately, as explained in \Sec{8BNi}, the larger Coulomb interaction leads the REB form factor to overestimate the DEA ratio.
It is therefore not clear that the gain in the breakup channel will improve the accuracy of the ratio method.

When the $p$-$T$ interaction is removed ($V_{pT}$=0, dashed red lines), the conclusions are very similar to those made for the $^{12}$C target.
For $^{27}$P and $^{17}$F the remnant oscillations in the ratio disappear and the ratio and its REB prediction are superimposed for $\theta>2^\circ$.
For $^{25}$Al, on the contrary, the ratio still exhibits significant remnant oscillations.
	
Results in \Fig{f10} also show that the potential $V_{cT}$ used to simulate the $c$-$T$ interaction has little to no influence on the cross sections, especially at small angles. This is to be expected from this more Coulomb-dominated process. However, at larger angles, at which some differences can be observed in the individual cross sections, the ratios are nearly unaffected by the choice of that interaction.

\begin{figure*}[htbp]
	\centering
	\makebox[\textwidth][c]{\includegraphics[width=1.35\textwidth]{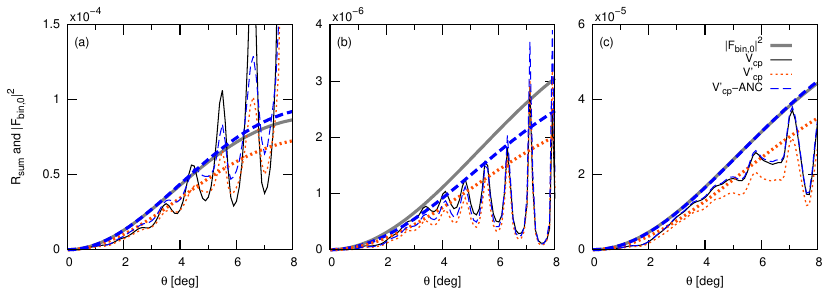}}
	\caption{Same as Fig.~\ref{f8} but on a \nuc{58}{Ni} target at 60~MeV/nucleon.}
	\label{f11}
\end{figure*}

To study the influence of the choice of $c$-$p$ potential on the ratio, we perform the same series of calculations as in Fig.~\ref{f8} for this $^{58}$Ni target; the results are displayed in Fig.~\ref{f11}.
Before analyzing the difference between the first $V_{cp}$ (black and grey solid lines) and the second $V'_{cp}$ (red dotted lines), let us first note that the agreement between the DEA ratio and its REB prediction is not as good as on $^{12}$C.
Although it leads to a larger---and hence easier to measure---breakup cross section, this choice of target is not fit for a \emph{strict} application of the ratio, even for $^{27}$P.
On a nickel target, the ratio could at best be used in its \emph{approximate} version.

The sensitivity of the method to the projectile description is very similar to what has been observed in \Fig{f8} on a $^{12}$C target.
In the case of $^{27}$P, the choice of the $c$-$p$ potential does not affect the quality of the method: the agreement  of the REB prediction with the DEA ratio does not vary much when $V'_{cp}$ is used instead of the original $V_{cp}$. That change is solely due to the difference in the ground-state ANC, as proven by the blue dashed lines, which display the $V'_{cp}$ results scaled to the ANC obtained with $V_{cp}$ (see \Sec{FAlPC}).
In the case of $^{17}$F, we observe again that the dynamical calculation of the reaction is less peripheral than predicted by the REB, the DEA ratio obtained with $V'_{cp}$ being not superimposed on the $V_{cp}$ calculation after its scaling by the ANC.
For $^{25}$Al, both the DEA and REB models indicate that the reaction is not peripheral.
Since the DEA calculation deviates so much from its REB estimate, only estimates of $l_0$ and/or $E_0$ could be gathered through the use of the \emph{approximate} version of the method.

Finally, it should be mentioned that we have also calculated the ratios ${\mathcal R}_{\rm sum}$ for several continuum bins.
We do not show these results for the sake of conciseness since they are similar to the carbon-target case: broadening the bin leads to larger breakup cross sections with a quality of the agreement between the DEA calculation and the REB prediction that remains poor for $^{17}$F and $^{25}$Al and that worsens at high energy for $^{27}$P.

\section{\label{sec:conclusion}Conclusion}

The ratio observable is a recent tool proposed in Refs.~\cite{capel2011,capel2013} to study the structure of loosely-bound systems like halo nuclei.
It consists of the ratio of elastic-scattering and breakup angular distributions and captures the projectile structure while showing little dependence on the reaction mechanism.
This method has been shown to work quite well for one-neutron halo nuclei at both intermediate \cite{capel2011,capel2013} and low \cite{colomer2016} beam energies.

In this work, we investigate the extension of this new reaction observable to study the single-particle structure of proton-rich nuclei.
To this end, we first study the ratio method for a $^{8}$B projectile, which has a clear $^7$Be-$p$ structure and is usually seen as the archetype of a charged halo nucleus.
We consider light (C), medium-mass (Ni) and heavy (Pb) targets at an intermediate beam energy of 44~MeV/nucleon.
To obtain a reliable estimate of the actual ratio, we use the DEA \cite{baye2005,goldstein2006}, which provides excellent agreement with experimental data for one-neutron \cite{goldstein2006} and one-proton \cite{GCB07} halo nuclei at these beam energies.
This enables us to study the validity of the method near the proton dripline and to study its sensitivity to the projectile structure.
We have then extended our study to other proton-rich nuclei, namely \ex{17}{F}, \ex{25}{Al}, and \ex{27}{P}.

Although the REB assumptions are not totally fulfilled (the Coulomb interaction between the valence proton and the target can hardly be neglected), we have shown that the ratio method applied to the proton-rich side of the nuclear chart still removes part of the dependence on the reaction mechanism, especially on the $c$-$T$ optical potential choice.
Unfortunately, the results obtained are less good than expected in the \emph{strict application} of the method developed for one-neutron halo nuclei in Refs.~\cite{capel2011,capel2013,colomer2016}; we observe that even for nuclei with a valence proton loosely bound in an $s$ or $p$ orbital, the DEA ratio exhibits remnant oscillations, which are mainly due to the proton-target interaction.
Nevertheless in these cases, it still follows the form factor predicted by the REB and a direct comparison of that form factor with actual measurements would be possible.
It would then provide a good estimate of the one-proton separation energy, its orbital angular momentum and the ANC of the ground-state wave function.
This is what we have coined the \emph{strict application} of the method.

For more deeply bound systems or when the valence proton sits in an $l\geq2$ orbital the original idea of the ratio lapses. The ratio method could however be applied \emph{approximately} thanks to the high sensitivity of the ratio to the one-proton separation energy and to its orbital angular momentum. This would enable us to infer valuable information about the structure of the nucleus. 
Alternatively, for such nuclei, experimental data could be compared to dynamical calculations of the ratio, e.g., using the DEA.
Although this \emph{dynamical} version of the ratio method does not exhibit the simplicity and elegance of the original ratio idea, it strongly reduces the dependence on the reaction mechanism and hence is more sensitive to details of the nuclear structure than single cross sections for scattering or breakup.

Similarly to what has been observed at low energies for one-neutron halo nuclei \cite{capel2011,capel2013,colomer2016}, the ratio is independent of the choice of optical potential used to simulate the interaction between the core of the nucleus and the target.
This in itself is very valuable since this interaction is usually poorly known and induces biases in the analysis of reaction data.
However, as seen in \Ref{colomer2016}, we find that the agreement with the REB is improved when the role of the Coulomb interaction is reduced.
This suggests that a direct comparison of an experimentally measured ratio with the REB form factor would be best performed on light targets.
However, breakup cross sections on such targets are small compared to heavier ones, on which the Coulomb breakup becomes significant.
The use of a broad energy bin in the $c$-$p$ continuum instead of a single energy enhances the ratio by several orders of magnitudes without much accuracy loss.

Most of the disagreement observed between the REB prediction and the DEA calculations arises from the $p$-$T$ potential, and in particular to its Coulomb part. In a near future, we plan to develop a correction to the REB to include this interaction at the perturbation level~\cite{Johnson-private}. We hope that with this correction, the REB form factor will account for most of the remnant oscillations we have observed in the present study. This should then significantly improve the ratio method on the proton-rich side of the nuclear chart.

This analysis shows that the applicability of the ratio method, which was originally designed for neutron-halo nuclei, can be extended to the proton-rich side of the nuclear chart under strict conditions, viz. the system must be loosely-bound and the valence proton must sit in a low-$l$ orbital. This interesting results shows the limitations of the ratio method in its original form. However, being less sensitive than usual reaction observables to the reaction mechanism and the particulars of the optical potentials and being very sensitive to the binding energy of such systems, the ratio offers an alternative to conventional reaction methods to study the structure of proton-rich nuclei.

\ack This work is supported by the National Natural Science Foundation of China (Grants No. 11775013 and U1432247) and the national key research and development program (2016YFA0400502), and the Research Credit No. 19526092 of the Belgian Funds for Scientific Research F.R.S.-FNRS. F.~C. acknowledges the support of the Deutsche Forschungsgemeinschaft (DFG) in the framework of the PRISMA cluster of excellence ``Precision Physics, Fundamental Interactions and structure of Matter''. This project has received funding from the European Union's Horizon 2020 research and innovation programme under grant agreement No 654002. This text presents research results of the Belgian Research Initiative on eXotic nuclei (BriX), Program No.~P7/12 on interuniversity attraction poles of the Belgian Federal Science Policy Office.

\section*{References}
\bibliographystyle{iopart-num}
\bibliography{paper8B}

\providecommand{\noopsort}[1]{}\providecommand{\singleletter}[1]{#1}%
\providecommand{\newblock}{}
\begin{thebibliography}{10}
\expandafter\ifx\csname url\endcsname\relax
  \def\url#1{{\tt #1}}\fi
\expandafter\ifx\csname urlprefix\endcsname\relax\def\urlprefix{URL }\fi
\providecommand{\eprint}[2][]{\url{#2}}

\bibitem{Tan85r}
Tanihata I, Hamagaki H, Hashimoto O, Shida Y, Yoshikawa N, Sugimoto K, Yamakawa
  O, Kobayashi T and Takahashi N 1985 {\em Phys. Rev. Lett.\/} {\bf 55}(24)
  2676--2679
  \urlprefix\url{https://link.aps.org/doi/10.1103/PhysRevLett.55.2676}

\bibitem{JRF04}
Jensen A~S, Riisager K, Fedorov D~V and Garrido E 2004 {\em Rev. Mod. Phys.\/}
  {\bf 76}(1) 215--261
  \urlprefix\url{https://link.aps.org/doi/10.1103/RevModPhys.76.215}

\bibitem{KHH16}
Kanungo R, Horiuchi W, Hagen G, Jansen G~R, Navratil P, Ameil F, Atkinson J,
  Ayyad Y, Cortina-Gil D, Dillmann I, Estrad\'e A, Evdokimov A, Farinon F,
  Geissel H, Guastalla G, Janik R, Kimura M, Kn\"obel R, Kurcewicz J, Litvinov
  Y~A, Marta M, Mostazo M, Mukha I, Nociforo C, Ong H~J, Pietri S, Prochazka A,
  Scheidenberger C, Sitar B, Strmen P, Suzuki Y, Takechi M, Tanaka J, Tanihata
  I, Terashima S, Vargas J, Weick H and Winfield J~S 2016 {\em Phys. Rev.
  Lett.\/} {\bf 117}(10) 102501
  \urlprefix\url{https://link.aps.org/doi/10.1103/PhysRevLett.117.102501}

\bibitem{OKS00}
Ozawa A, Kobayashi T, Suzuki T, Yoshida K and Tanihata I 2000 {\em Phys. Rev.
  Lett.\/} {\bf 84}(24) 5493--5495
  \urlprefix\url{https://link.aps.org/doi/10.1103/PhysRevLett.84.5493}

\bibitem{KNP09}
Kanungo R, Nociforo C, Prochazka A, Aumann T, Boutin D, Cortina-Gil D, Davids
  B, Diakaki M, Farinon F, Geissel H, Gernh\"auser R, Gerl J, Janik R, Jonson
  B, Kindler B, Kn\"obel R, Kr\"ucken R, Lantz M, Lenske H, Litvinov Y, Lommel
  B, Mahata K, Maierbeck P, Musumarra A, Nilsson T, Otsuka T, Perro C,
  Scheidenberger C, Sitar B, Strmen P, Sun B, Szarka I, Tanihata I, Utsuno Y,
  Weick H and Winkler M 2009 {\em Phys. Rev. Lett.\/} {\bf 102}(15) 152501
  \urlprefix\url{https://link.aps.org/doi/10.1103/PhysRevLett.102.152501}

\bibitem{STA13}
Steppenbeck D, Takeuchi S, Aoi N, Doornenbal P, Matsushita M, Wang H, Baba H,
  Fukuda N, Go S, Honma M, Lee J, Matsui K, Michimasa S, Motobayashi T,
  Nishimura D, Otsuka T, Sakurai H, Shiga Y, Söderström P~A, Sumikama T,
  Suzuki H, Taniuchi R, Utsuno Y, Valiente-Dobón J~J and Yoneda K 2013 {\em
  Nature\/} {\bf 502} 207 \urlprefix\url{http://dx.doi.org/10.1038/nature12522}

\bibitem{RAA15}
Rosenbusch M, Ascher P, Atanasov D, Barbieri C, Beck D, Blaum K, Borgmann C,
  Breitenfeldt M, Cakirli R~B, Cipollone A, George S, Herfurth F, Kowalska M,
  Kreim S, Lunney D, Manea V, Navr\'atil P, Neidherr D, Schweikhard L, Som\`a
  V, Stanja J, Wienholtz F, Wolf R~N and Zuber K 2015 {\em Phys. Rev. Lett.\/}
  {\bf 114}(20) 202501
  \urlprefix\url{https://link.aps.org/doi/10.1103/PhysRevLett.114.202501}

\bibitem{HW11}
Heyde K and Wood J~L 2011 {\em Rev. Mod. Phys.\/} {\bf 83}(4) 1467--1521
  \urlprefix\url{https://link.aps.org/doi/10.1103/RevModPhys.83.1467}

\bibitem{YWX16}
Yang X~F, Wraith C, Xie L, Babcock C, Billowes J, Bissell M~L, Blaum K, Cheal
  B, Flanagan K~T, Garcia~Ruiz R~F, Gins W, Gorges C, Grob L~K, Heylen H,
  Kaufmann S, Kowalska M, Kraemer J, Malbrunot-Ettenauer S, Neugart R, Neyens
  G, N\"ortersh\"auser W, Papuga J, S\'anchez R and Yordanov D~T 2016 {\em
  Phys. Rev. Lett.\/} {\bf 116}(18) 182502
  \urlprefix\url{https://link.aps.org/doi/10.1103/PhysRevLett.116.182502}

\bibitem{Tan96}
Tanihata I 1996 {\em J. Phys. G\/} {\bf 22} 157--198
  \urlprefix\url{http://stacks.iop.org/0954-3899/22/i=2/a=004}

\bibitem{BC12}
Baye D and Capel P 2012 Breakup reaction models for two- and three-cluster
  projectiles {\em Clusters in Nuclei, Vol. 2\/} vol 848 ed Beck C (Heidelberg:
  Springer)

\bibitem{CGB04}
Capel P, Goldstein G and Baye D 2004 {\em Phys. Rev. C\/} {\bf 70}(6) 064605
  \urlprefix\url{https://link.aps.org/doi/10.1103/PhysRevC.70.064605}

\bibitem{capel2011}
Capel P, Johnson R~C and Nunes F~M 2011 {\em Phys. Lett.\/} {\bf B705} 112
  \urlprefix\url{http://www.sciencedirect.com/science/article/pii/S0370269311012056}

\bibitem{capel2013}
Capel P, Johnson R~C and Nunes F~M 2013 {\em Phys. Rev. C\/} {\bf 88} 044602
  \urlprefix\url{http://link.aps.org/doi/10.1103/PhysRevC.88.044602}

\bibitem{johnson1997}
Johnson R~C, Al-Khalili J~S and Tostevin J~A 1997 {\em Phys. Rev. Lett\/} {\bf
  79} 2771 \urlprefix\url{http://link.aps.org/doi/10.1103/PhysRevLett.79.2771}

\bibitem{RCJ1997}
Johnson R~C 1999 Elastic scattering and breakup of halo nuclei in a special
  model {\em Proc. of the Euro. Conf. in Advances in Nucl. Phys. and Related
  Areas {\rm (July 1997, Thessaloniki, Greece)}\/} ed Brink D, Grypeos M and
  Massen S (Thessaloniki: Giahoudi-Giapouli Publishing) p 156

\bibitem{colomer2016}
Colomer F, Capel P, Nunes F~M and Johnson R~C 2016 {\em Phys. Rev. C\/} {\bf
  93} 054621
  \urlprefix\url{https://link.aps.org/doi/10.1103/PhysRevC.93.054621}

\bibitem{baye2005}
Baye D, Capel P and Goldstein G 2005 {\em Phys. Rev. Lett\/} {\bf 95} 082502
  \urlprefix\url{http://link.aps.org/doi/10.1103/PhysRevLett.95.082502}

\bibitem{goldstein2006}
Goldstein G, Baye D and Capel P 2006 {\em Phys. Rev. C\/} {\bf 73} 024602
  \urlprefix\url{https://link.aps.org/doi/10.1103/PhysRevC.73.024602}

\bibitem{GCB07}
Goldstein G, Capel P and Baye D 2007 {\em Phys. Rev. C\/} {\bf 76}(2) 024608
  \urlprefix\url{https://link.aps.org/doi/10.1103/PhysRevC.76.024608}

\bibitem{Togano-PRC-2011}
Togano Y, Gomi T, Motobayashi T, Ando Y, Aoi N, Baba H, Demichi K, Elekes Z,
  Fukuda N, F\"ul\"op Z, Futakami U, Hasegawa H, Higurashi Y, Ieki K, Imai N,
  Ishihara M, Ishikawa K, Iwasa N, Iwasaki H, Kanno S, Kondo Y, Kubo T, Kubono
  S, Kunibu M, Kurita K, Matsuyama Y~U, Michimasa S, Minemura T, Miura M,
  Murakami H, Nakamura T, Notani M, Ota S, Saito A, Sakurai H, Serata M,
  Shimoura S, Sugimoto T, Takeshita E, Takeuchi S, Ue K, Yamada K, Yanagisawa
  Y, Yoneda K and Yoshida A 2011 {\em Phys. Rev. C\/} {\bf 84}(3) 035808
  \urlprefix\url{https://link.aps.org/doi/10.1103/PhysRevC.84.035808}

\bibitem{Chen-PRC-2012}
Chen J, Chen A~A, Am\'adio G, Cherubini S, Fujikawa H, Hayakawa S, He J~J,
  Iwasa N, Kahl D, Khiem L~H, Kubono S, Kurihara S, Kwon Y~K, La~Cognata M,
  Moon J~Y, Niikura M, Nishimura S, Pearson J, Pizzone R~G, Teranishi T, Togano
  Y, Wakabayashi Y and Yamaguchi H 2012 {\em Phys. Rev. C\/} {\bf 85}(1) 015805
  \urlprefix\url{https://link.aps.org/doi/10.1103/PhysRevC.85.015805}

\bibitem{Jung-PRC-2012}
Jung H~S, Lee C~S, Kwon Y~K, Moon J~Y, Lee J~H, Yun C~C, Kubono S, Yamaguchi H,
  Hashimoto T, Kahl D, Hayakawa S, Choi S, Kim M~J, Kim Y~H, Kim Y~K, Park J~S,
  Kim E~J, Moon C~B, Teranishi T, Wakabayashi Y, Iwasa N, Yamada T, Togano Y,
  Kato S, Cherubini S and Rapisarda G~G 2012 {\em Phys. Rev. C\/} {\bf 85}(4)
  045802 \urlprefix\url{https://link.aps.org/doi/10.1103/PhysRevC.85.045802}

\bibitem{Fortune-PRC-2015}
Fortune H~T 2015 {\em Phys. Rev. C\/} {\bf 92}(2) 025807
  \urlprefix\url{https://link.aps.org/doi/10.1103/PhysRevC.92.025807}

\bibitem{Marganiec-PRC-2016}
Marganiec J, Beceiro~Novo S, Typel S, Langer C, Wimmer C, Alvarez-Pol H, Aumann
  T, Boretzky K, Casarejos E, Chatillon A, Cortina-Gil D, Datta-Pramanik U,
  Elekes Z, Fulop Z, Galaviz D, Geissel H, Giron S, Greife U, Hammache F, Heil
  M, Hoffman J, Johansson H, Kiselev O, Kurz N, Larsson K, Le~Bleis T, Litvinov
  Y~A, Mahata K, Muentz C, Nociforo C, Ott W, Paschalis S, Plag R, Prokopowicz
  W, Rodr\'{\i}guez~Tajes C, Rossi D~M, Simon H, Stanoiu M, Stroth J,
  S\"ummerer K, Wagner A, Wamers F, Weick H and Wiescher M (R3B Collaboration)
  2016 {\em Phys. Rev. C\/} {\bf 93}(4) 045811
  \urlprefix\url{https://link.aps.org/doi/10.1103/PhysRevC.93.045811}

\bibitem{Hag10}
Hagen G, Papenbrock T and Hjorth-Jensen M 2010 {\em Phys. Rev. Lett.\/} {\bf
  104}(18) 182501
  \urlprefix\url{http://link.aps.org/doi/10.1103/PhysRevLett.104.182501}

\bibitem{XuXX-PLB-2013}
Xu X, Lin C, Jia H, Yang F, Zhang H, Liu Z, Wu Z, Yang L, Bao P, Sun L, Xu H,
  Wang J, Yang Y, Sun Z, Hu Z, Wang M, Jin S, Han J, Zhang N, Chen S, Lei X,
  Huang M, Ma P, Ma J, Zhang Y, Zhou X, Ma X and Xiao G 2013 {\em Phys.
  Lett.\/} {\bf B727} 126 -- 129 ISSN 0370-2693
  \urlprefix\url{http://www.sciencedirect.com/science/article/pii/S0370269313008356}

\bibitem{XP13}
Xu Y~P and Pang D~Y 2013 {\em Phys. Rev. C\/} {\bf 87}(4) 044605
  \urlprefix\url{https://link.aps.org/doi/10.1103/PhysRevC.87.044605}

\bibitem{CEN12}
Capel P, Esbensen H and Nunes F~M 2012 {\em Phys. Rev. C\/} {\bf 85}(4) 044604
  \urlprefix\url{https://link.aps.org/doi/10.1103/PhysRevC.85.044604}

\bibitem{Dav98}
Davids B, Anthony D~W, Austin S~M, Bazin D, Blank B, Caggiano J~A, Chartier M,
  Esbensen H, Hui P, Powell C~F, Scheit H, Sherrill B~M, Steiner M, Thirolf P,
  Yurkon J and Zeller A 1998 {\em Phys. Rev. Lett.\/} {\bf 81} 2209

\bibitem{Dav01l}
Davids B, Anthony D~W, Aumann T, Austin S~M, Baumann T, Bazin D, Clement R~R~C,
  Davids C~N, Esbensen H, Lofy P~A, Nakamura T, Sherrill B~M and Yurkon J 2001
  {\em Phys. Rev. Lett.\/} {\bf 86} 2750

\bibitem{Dav01c}
Davids B, Austin S~M, Bazin D, Esbensen H, Sherrill B~M, Thompson I~J and
  Tostevin J~A 2001 {\em Phys. Rev. C\/} {\bf 63} 065806

\bibitem{fukui2014}
Fukui T, Ogata K and Capel P 2014 {\em Phys. Rev. C\/} {\bf 90} 034617
  \urlprefix\url{http://link.aps.org/doi/10.1103/PhysRevC.90.034617}

\bibitem{BDT94}
Baye D, Descouvemont P and Timofeyuk N 1994 {\em Nucl. Phys. A\/} {\bf 577} 624
  -- 640 ISSN 0375-9474
  \urlprefix\url{http://www.sciencedirect.com/science/article/pii/0375947494909369}

\bibitem{MTT02}
Mortimer J, Thompson I~J and Tostevin J~A 2002 {\em Phys. Rev. C\/} {\bf 65}
  064619

\bibitem{esbensen1996}
Esbensen H and Bertsch G~F 1996 {\em Nucl. Phys. A\/} {\bf 600} 37
  \urlprefix\url{http://www.sciencedirect.com/science/article/pii/0375947496000061}

\bibitem{cook1982}
Cook J 1982 {\em Nucl. Phys; A\/} {\bf 388} 153--172
  \urlprefix\url{http://www.sciencedirect.com/science/article/pii/0375947482905139}

\bibitem{nadasen1995}
Nadasen A, Brusoe J, Farhat J, Stevens T, Williams J, Nieman L, Winfield J~S,
  Warner R~E, Becchetti F~D, J\"anecke J~W, Annakkage T, Bajema J, Roberts D
  and Govinden H~S 1995 {\em Phys. Rev. C\/} {\bf 52} 1894--1899
  \urlprefix\url{https://link.aps.org/doi/10.1103/PhysRevC.52.1894}

\bibitem{CLARK1995416}
Clark H, Lui Y~W and Youngblood D 1995 {\em Nucl. Phys. A\/} {\bf 589} 416
  \urlprefix\url{http://www.sciencedirect.com/science/article/pii/037594749500121G}

\bibitem{chuev1971}
Chuev V~I, Davidov V~V, Novatskii B~G, Ogloblin A~A, Sakuta S~B and Stepanov
  D~N 1971 {\em J. Phys. Colloques\/} {\bf 32} 157
  \urlprefix\url{https://jphyscol.journaldephysique.org/articles/jphyscol/abs/1971/06/jphyscol197132C626/jphyscol197132C626.html}

\bibitem{schumacher1973}
Schumacher P, Ueta N, Duhm H~H, Kubo K~I and Klages W~J 1973 {\em Nucl. Phys.
  A\/} {\bf 212} 573
  \urlprefix\url{http://www.sciencedirect.com/science/article/pii/0375947473908245?via%3Dihub}

\bibitem{perey1976}
Perey C~M and Perey F~G 1976 {\em At. Data Nucl. Data Tables\/} {\bf 17} 1--101
  \urlprefix\url{http://dx.doi.org/10.1016/0092-640X(76)90007-3}

\bibitem{becchetti1969}
Becchetti F~D and Greenlees G~W 1969 {\em Phys. Rev.\/} {\bf 182}(4) 1190--1209
  \urlprefix\url{https://link.aps.org/doi/10.1103/PhysRev.182.1190}

\bibitem{koning2003}
Koning A~J and Delaroche J~P 2003 {\em Nucl. Phys. A\/} {\bf 713} 231--310
  \urlprefix\url{http://www.sciencedirect.com/science/article/pii/S0375947402013210}

\bibitem{capel2003}
Capel P, Baye D and Melezhik V~S 2003 {\em Phys. Rev. C\/} {\bf 68} 014612
  \urlprefix\url{https://link.aps.org/doi/10.1103/PhysRevC.68.014612}

\bibitem{capel2010}
Capel P, Hussein M and Baye D 2010 {\em Phys. Lett.\/} {\bf B693} 448
  \urlprefix\url{http://www.sciencedirect.com/science/article/pii/S0370269310010452}

\bibitem{Johnson-private}
Johnson R~C 2017 (private communication)

\bibitem{Barker}
{Barker} F~C 1980 {\em Australian Journal of Physics\/} {\bf 33} 177--190
  \urlprefix\url{http://adsabs.harvard.edu/abs/1980AuJPh..33..177B}

\bibitem{Fuk04}
Fukuda N, Nakamura T, Aoi N, Imai N, Ishihara M, Kobayashi T, Iwasaki H, Kubo
  T, Mengoni A, Notani M, Otsu H, Sakurai H, Shimoura S, Teranishi T, Watanabe
  Y~X and Yoneda K 2004 {\em Phys. Rev. C\/} {\bf 70}(5) 054606
  \urlprefix\url{https://link.aps.org/doi/10.1103/PhysRevC.70.054606}

\bibitem{SBI00}
Sparenberg J~M, Baye D and Imanishi B 2000 {\em Phys. Rev. C\/} {\bf 61} 054610

\bibitem{Firestone-NDS-2009}
Firestone R 2009 {\em Nucl. Data Sheets\/} {\bf 110} 1691 -- 1744 ISSN
  0090-3752
  \urlprefix\url{http://www.sciencedirect.com/science/article/pii/S009037520900057X}

\bibitem{Basunia-NDS-2011}
Basunia M~S 2011 {\em Nucl. Data Sheets\/} {\bf 112} 1875 -- 1948 ISSN
  0090-3752
  \urlprefix\url{http://www.sciencedirect.com/science/article/pii/S0090375211000664}

\bibitem{CH89}
Varner R, Thompson W, McAbee T, Ludwig E and Clegg T 1991 {\em Phys. Rep.\/}
  {\bf 201} 57 -- 119 ISSN 0370-1573
  \urlprefix\url{http://www.sciencedirect.com/science/article/pii/037015739190039O}

\bibitem{Bauge-PRC-2001}
Bauge E, Delaroche J~P and Girod M 2001 {\em Phys. Rev. C\/} {\bf 63}(2) 024607
  \urlprefix\url{https://link.aps.org/doi/10.1103/PhysRevC.63.024607}

\bibitem{Brown-PRC-1998}
Alex~Brown B 1998 {\em Phys. Rev. C\/} {\bf 58}(1) 220--231
  \urlprefix\url{https://link.aps.org/doi/10.1103/PhysRevC.58.220}

\end{thebibliography}

\end{document}